\documentclass[twocolumn]{aastex631}
\usepackage{amsmath}
\usepackage{CJK}
\newcommand{\mtot}{$M_{\rm tot}$}
\newcommand{\msun}{$M_{\odot}$}
\newcommand{\mpcyr}{$\:\rm{Mpc^{-3}\:yr^{-1}}$}

\begin{document}
\begin{CJK*}{UTF8}{gbsn}

\title{Contaminating Electromagnetic Transients in LISA Gravitational \\Wave Localization Volumes. I: The Intrinsic Rates}

\correspondingauthor{Weixiang Yu}
\email{wyu@ubishops.ca}

\author[0000-0003-1262-2897]{Weixiang Yu~(于伟翔)}
\affil{Department of Physics \& Astronomy, Bishop's University, 2600 rue College, Sherbrooke, QC, J1M 1Z7, Canada}

\author[0000-0001-8665-5523]{John~J.~Ruan}
\affil{Department of Physics \& Astronomy, Bishop's University, 2600 rue College, Sherbrooke, QC, J1M 1Z7, Canada}

\author[0000-0002-3719-940X]{Michael Eracleous}
\affil{Department of Astronomy \& Astrophysics and Institute for Gravitation and the Cosmos, Penn State University, 525 Davey Lab, 251
Pollock Road, University Park, PA 16802, USA}

\author[0000-0001-8557-2822]{Jessie Runnoe}
\affil{Department of Physics \& Astronomy, Vanderbilt University, 2301 Vanderbilt Place, Nashville, TN 37235, USA}

\author[0000-0001-6803-2138]{Daryl Haggard}
\affil{Trottier Space Institute \& Department of Physics, McGill University, 3600 rue University, Montreal, QC, H3A 2T8, Canada}

\author[0000-0002-7835-7814]{Tamara Bogdanovi\'c}
\affil{School of Physics \& Center for Relativistic Astrophysics, 837 State St NW, Georgia Institute of Technology, Atlanta, GA 30332, USA}

\author[0000-0003-1542-0799]{Aaron Stemo}
\affil{Department of Physics \& Astronomy, Vanderbilt University, 2301 Vanderbilt Place, Nashville, TN 37235, USA}

\author[0009-0000-4083-2547]{Kaitlyn Szekerczes}
\affil{Department of Astronomy \& Astrophysics and Institute for Gravitation and the Cosmos, Penn State University, 525 Davey Lab, 251
Pollock Road, University Park, PA 16802, USA}

\author[0009-0006-1022-5627]{Carolyn L. Drake}
\affil{Department of Physics \& Astronomy, Vanderbilt University, 2301 Vanderbilt Place, Nashville, TN 37235, USA}

\author[0000-0002-5456-9134]{Kate E. Futrowsky}
\affil{School of Physics \& Center for Relativistic Astrophysics, 837 State St NW, Georgia Institute of Technology, Atlanta, GA 30332, USA}

\author[0000-0002-8187-1144]{Steinn Sigurdsson}
\affil{Department of Astronomy \& Astrophysics and Institute for Gravitation and the Cosmos, Penn State University, 525 Davey Lab, 251
Pollock Road, University Park, PA 16802, USA}

\begin{abstract}
The Laser Interferometer Space Antenna (LISA) will soon detect gravitational waves (GWs) emitted by massive black hole (MBH) mergers. Some theoretical models have predicted transient electromagnetic (EM) emission from these mergers, enabling the association of LISA GW sources with their EM counterparts via telescope follow-up. However, the number of unrelated EM transients that might contaminate telescope searches for the true transient counterparts of LISA MBH mergers is unknown. 
We investigate the expected numbers of unrelated EM transients that will coincide with simulated LISA localization volumes of MBH mergers, as a function of the merger total mass and redshift. 
We find that the number of potential contaminants in LISA localization volumes drops to unity for mergers at $z \lesssim 0.8$ and at 1 hour before coalescence. After coalescence, the parameter space corresponding to a maximum of one potential contaminant expands to $z \lesssim 1.5$. In contrast, if the redshifts for all transients detected in LISA sky localization regions are not available, the number of potential contaminants increases by an average factor of $\sim100$, and never drops below unity.
Overall, we expect the average number of contaminating transients in telescope follow-up of LISA MBH mergers to be non-negligible, especially without redshift information for the detected transients. We recommend that endeavors designing follow-up strategies of LISA events should focus on: (1) building large redshift catalogs for host galaxies, (2) developing robust real-time transient classification algorithms, (3) and coordinating telescope resources to obtain redshifts for candidate transient EM counterparts in a timely manner.

\end{abstract}
\keywords{Gravitational waves -- Massive black holes -- Transient sources -- Optical counterparts}

\section{Introduction} \label{sec:bkg}
In hierarchical galaxy formation, massive black hole (MBH) binaries are natural products of galaxy mergers~\citep{Volonteri2003, Hopkins2008}.
Massive galaxies are not born massive, but grow primarily through sequences of mergers with other galaxies~\citep{Fakhouri2010, Rodriguez-Gomez2015, Behroozi2019}. 
It is believed that nearly all galaxies host MBHs ($\gtrsim10^5\,$\msun) at their centers~\citep{Kormendy2013}. 
When two galaxies merge, their MBHs are expected to sink toward their new common gravitational potential center through dynamical friction~\citep{Chandrasekhar1943}, and form a bound MBH binary~\citep{Begelman1980, Escala2004, Dotti2006}. 
This MBH binary hardens through stellar scattering~\citep[e.g.,][]{Quinlan1996, Khan2013, Sesana2015, Khan2020} and gas torques~\citep[e.g.,][]{Escala2004, Cuadra2009, Duffell2020}, until the angular momentum loss is dominated by gravitational wave (GW) emission~\citep{Peters1964}. The MBH binary will eventually coalesce, and the resultant GW chirp will be detected by future low-frequency GW experiments~\citep[e.g.,][]{Wyithe2003, Sesana2004}. 

The upcoming Laser Interferometer Space Antenna~\citep[LISA;][]{Amaro-Seoane2017, colpi2024} will detect GWs emitted by MBH mergers.
LISA is sensitive to GWs in the 0.1 - 100~$\rm mHz$ frequency range, which are emitted by coalescing massive black hole binaries (MBHBs) of mass~$10^4 - 10^8$~\msun~up to a redshift of 20~\citep{colpi2024}. 
For a subset of these merging MBHBs, LISA can detect the GWs weeks before coalescence, depending on their total mass and redshift. 
By detecting samples of MBH mergers with a range of intrinsic properties (e.g., total mass, mass ratio, redshift, spin, etc.), LISA will probe the origins and the evolution of MBHs over cosmic time, as well as unveil the extreme environments in the vicinity of coalescing MBHBs~\citep{Amaro-Seoane2023, colpi2024}. 

Telescope follow-up of MBH mergers detected by LISA will enable a range of multi-messenger science investigations. 
For example, pre-merger monitoring of coalescing MBHBs across the electromagnetic (EM) spectrum will open a unique window to examine how the surrounding gas reacts to the rapidly changing strong gravitational potential~\citep{Armitage2002, Amaro-Seoane2023}.
The GW waveform encodes information on the fundamental properties (e.g., black hole mass and spin) of the MBHs in the binary, hence joint GW and EM observations of actively accreting MBHs in the binary may also allow us to test and validate EM-based methods to infer MBH fundamental properties~\citep[e.g.,][]{McClintock2011, Peterson2014}. 
Finally, MBH mergers detected in both GW and EM are bright standard sirens for GW cosmology~\citep{Holz2005, Abbott2017, Tamanini2017, Laghi2021, Mangiagli2023}. 
However, these science opportunities critically depend on our ability to confidently detect EM emission associated with the MBH merger in telescope follow-up observations.

A variety of transient time-domain EM signatures have been proposed to arise from MBH mergers.
During the last few orbits of the merger, prompt transient flashes in Poynting flux could arise due to winding of the magnetic fields by the binary~\citep{Palenzuela2010, Kelly2017}. 
If the accretion onto the MBHs in the binary is radiatively inefficient, the hot plasma in the accretion flow can be heated up to a temperature of $\sim10^{12}\,$K (e.g., by shocks) as the orbit shrinks; this hot plasma will be devoured by the final MBH immediately following the merger. 
The gradually rising temperature/luminosity of the flow preceding the merger and a sudden decline following the merger due to the disappearance of the hot plasma gives rise to the characteristic light curve of a transient~\citep{Bode2010, Bode2012, Farris2010, Bogdanovic2011}. 
Weeks to months after coalescence, delayed transients could be observable due to shocks in the accretion flows, as well as jet launching~\citep{Lippai2008, Schnittman2008, Rossi2010, Corrales2010, Zanotti2010, Ravi2018, Yuan2021}. 
However, despite this rich collection of predicted transient EM signatures, none are definitive, since MBH mergers are yet to be detected in GWs. 
Thus, telescope follow-up strategies to search for EM counterparts must consider a wide range of possibilities.

Transient phenomena are common in the extragalactic sky, and thus telescope follow-up of LISA localization regions should expect to detect a significant number of unrelated transients that will complicate secure identification of any true EM transient counterpart. 
For example, the upcoming Rubin Observatory Legacy Survey of Space and Time~\citep[LSST;][]{ivezic2019}, a $\sim$18,000 $\rm deg^2$ all-sky survey, is expected to detect $\sim10^5$ supernova explosions per year up to a redshift of 1~\citep{LSSTScienceCollaboration2009}. 
Nuclear transients such as tidal disruption events are also expected to occur for MBHs in the mass range that LISA is most sensitive to~\citep{Rees1988, colpi2024}. 
Given our ignorance of the actual transient EM signatures of LISA MBH mergers, the challenge of identifying the EM counterpart may be significant. 
Therefore, it is critical to quantify the number of unrelated EM transients that will occur in the LISA localization volume, as a first step to design effective strategies for follow-up searches for the true transient EM counterpart.

In this work, we aim to quantify the number of unrelated extragalactic EM transients that are expected to occur in LISA GW localization volumes of merging MBHBs, at different times before and after coalescence.
Assessing the number unrelated transients before the merger guides the design of telescope follow-up strategies that will probe the evolution of MBHBs and their accretion flows near coalescence, which is impossible with post-merger EM observations. 
We focus on quantifying the intrinsic rates of unrelated transients, and do not carefully consider the actual observing scenarios (e.g., the detectability of a particular transient given its peak luminosity and the limiting flux of the follow-up telescope used).
We assume that the observer obtains a single observation covering the entire LISA sky localization region, compares the new images taken with archival images obtained for the same region, and counts the number of newly-detected sources.
We carry out two complementary investigations. 
In the first, the simulated MBH mergers follow a prescribed distribution over redshift, which is predicted by the NewHorizon simulation~\citep{Volonteri2020}. 
The goal of this first investigation is to assess the average number of unrelated transients in LISA localization volumes as a function of time, before and after coalescence.  
In the second, we simulate MBH mergers on a grid of total mass ($M_{\rm tot}$) and redshift ($z$), and investigate how the projected number of unrelated transients vary as function of \mtot~and~$z$ at different times before and after coalescence. 
For each investigation, we also consider two extreme scenarios: in one scenario we assume that redshifts can be acquired for all transients detected in LISA sky localization regions, and in the other scenario we assume that redshifts for detected transients are not available. 
In a follow-up paper, we will incorporate considerations regarding the detectability of unrelated transients into our calculations, based on the exact light curve and spectral energy density evolution of the transients, as well as the telescopes used in the observations.

This paper is organized as follows: In Section~\ref{sec:method}, we detail our catalogs of simulated MBH merger events (Section~\ref{subsec:megerRates}), our method to generate LISA localization volumes for each MBH merger event (Section~\ref{subsec:errVol}), the classes of extragalactic transients considered (Section~\ref{subsec:commonTrans}), and how we calculate the number of unrelated transients given a LISA localization volume. In Section~\ref{sec:results}, we present the results of our two investigations. 
In Section~\ref{sec:discuss}, we discuss the implications of our results on associating MBH mergers detected by LISA with their transient EM counterparts, as well as possible improvements and extensions of our current approach. We briefly summarize and conclude in Section~\ref{sec:conclude}. Throughout this paper, we assume a standard flat $\Lambda$CDM cosmology with $H_0 = 70$~km~s$^{-1}$~Mpc$^{-1}$, $\Omega_{m,0} = 0.3$, and $\Omega_{\Lambda,0} = 0.7$.

\section{Methodology} \label{sec:method}

\subsection{Catalogs of Simulated MBH Mergers}\label{subsec:megerRates}
\begin{figure*}
    \centering
    \epsscale{1.16}
    \plotone{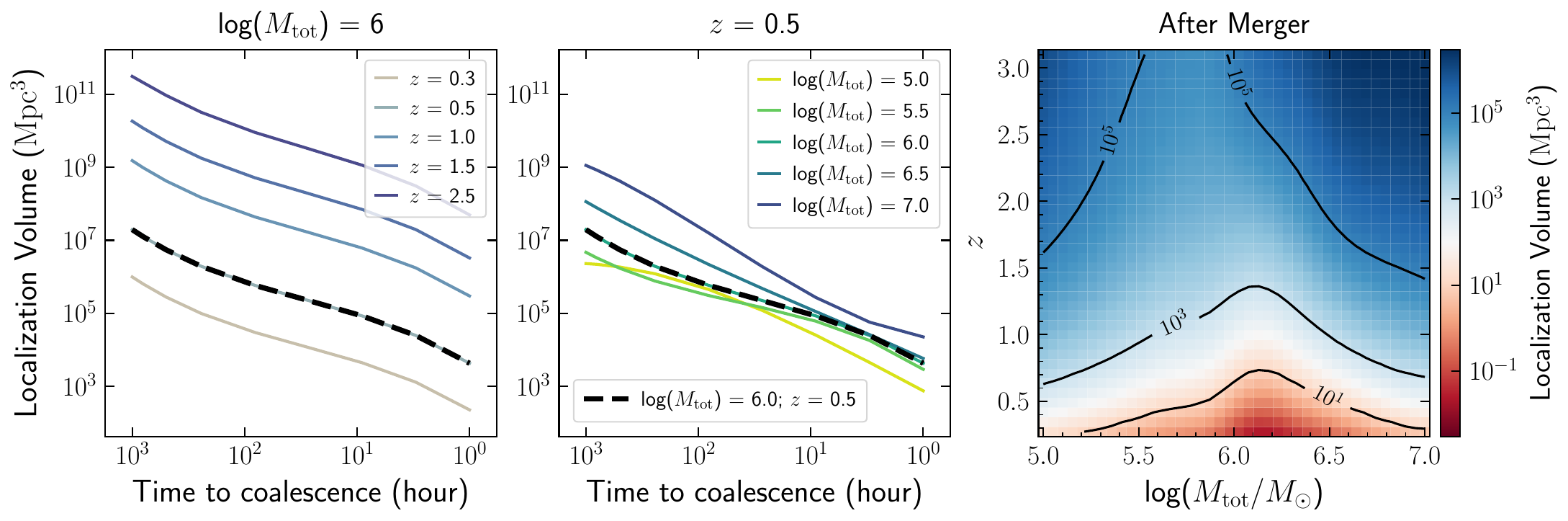}  
    \caption{LISA localization (comoving) volume as a function of MBH merger total mass (\mtot), redshift ($z$), and the time to coalescence. 
    The LISA localization volume increases with the merger mass and redshift and decreases with time to coalescence before the merger. After the merger, the localization volume is smallest for mergers with a \mtot~slightly greater than $10^{6}$\,\msun~at $z < 1$; the \mtot~corresponding to the smallest localization volume shifts towards $\sim10^{5.8}$\msun~at $z > 1$.
    \textit{Left}: \mtot~is fixed at $10^6$~\msun, and each curve presents the localization volumes of MBH mergers at a particular redshift. 
    The black dashed curve corresponds to MBH mergers at $z = 0.5$.
    \textit{Middle}: $z$~is fixed at 0.5, and each curve presents the localization volumes for MBH mergers with a particular~\mtot. 
    The black dashed curve corresponds to MBH mergers with~\mtot$=10^6\,$\msun.
    \textit{Right}: The dependence of the localization volume on merger total mass (x-axis) and redshift (y-axis) after coalescence.
    }
    \label{fig:errvol}
\end{figure*}

We create two catalogs of simulated MBH mergers, one for each of our two investigations, where each merger is uniquely defined by \mtot~and $z$.
The catalog used in our first investigation consists of 141,834 MBH mergers, simulated following the merger rate distribution (over $z$) produced by the NewHorizon simulation~\citep{Volonteri2020}. 
We note that the exact number of mergers simulated is not the focus of this investigation, since our goal is to compute the average number of contaminating EM transients.
{The NewHorizon simulation predicts a total of $\sim5$ mergers over the mission timespan of LISA, although \cite{Volonteri2020} emphasize that this is only a lower limit estimate.}
We adopt a flat distribution for the total mass of each merger from $10^{5}$\,\msun~to $10^{7}$\,\msun.
In other words, we set $dN/dM_{\rm tot} \propto\, (M_{\rm tot})^{\alpha}$, where $N$ is the number of mergers and $\alpha = 0$. 
This flat mass distribution is justified, because the local MBH mass function is nearly flat in our considered mass range~\citep[e.g.,][]{Greene2007}.
{In addition, the MBHs in the final sample of mergers in \cite{Volonteri2020}, after incorporating dynamical time delays in post-processing, also have a nearly flat mass distribution.}
In our tests, we find that our results do not change for a moderately different value of $\alpha$ (i.e., ${-0.4} < \alpha < 0.2$).
The second catalog, which is used in our second investigation, has a total of 1600 simulated MBH mergers.
The simulated MBH mergers in this catalog are drawn from an evenly-sampled 2D grid of total binary mass (spanning $M_{\rm tot} = 10^{5}$\,\msun~to $10^{7}$\,\msun) and redshift (spanning $z = 0.2$ to $3.2$).  
This mass and redshift range are chosen to enable us to quickly compute LISA localization volumes using the method of~\cite{mangiagli2020}. 

The mass ratio of simulated mergers is assumed to be randomly drawn from a range of 0.1 to 1.
We assume a flat distribution in mass ratio, because our adopted method to quickly compute LISA localization volumes does not take mass ratio as an input~\citep{mangiagli2020}. We will introduce the methodology of~\cite{mangiagli2020} and the associated underlying assumptions in Section~\ref{subsec:errVol}.
Future work will aim to investigate the dependence of the expected number of contaminating transients on the mass ratio of mergers.

\subsection{LISA Localization Volumes}\label{subsec:errVol}
The localization volume of a MBH merger detected by LISA can be estimated using the GW parameters extracted from the GW signal. 
The GW signal is determined by the fundamental properties of the MBHB~\citep{colpi2024}, including the luminosity distance ($d_{\rm L}$), the location on the sky ($\Omega$), and the time until coalescence ($t_{\rm c}$). These parameters can be inferred by fitting a waveform model to the observed GW signal~\citep[e.g.,][]{Santamaria2010, Klein2014, London2018}.
Given $d_{\rm L}$ and $\Omega$, along with their uncertainties $\Delta d_{\rm L}$ and $\Delta \Omega$, the localization volume can be approximated as the integrated comoving volume between $d_{\rm L} - \Delta d_{\rm L}$ and $d_{\rm L} + \Delta d_{\rm L}$ projected onto to the sky localization region $\Delta \Omega$.
More precisely, this is:
\begin{equation}\label{eqn:vol}
    V_{\rm loc} = \frac{\Delta \Omega}{4\pi}\int_{z_{\min}}^{z_{\max}} {dV_{\rm c}(z)}\,dz,
\end{equation}
where $z_{\min}$ is the redshift at the luminosity distance of  $d_{\rm L} - \Delta d_{\rm L}$ and $z_{\max}$ is the redshift at $d_{\rm L} + \Delta d_{\rm L}$. $dV_{\rm c}(z)$ is the infinitesimal comoving volume for an infinitesimal $dz$ at $z$.

To compute the localization volume for our simulated MBH mergers, we adopt the method introduced in \cite{mangiagli2020} to quickly compute the GW parameters of a given merger.
A total of 15 parameters are needed to fully describe the GW signal of a MBHB in a quasi-circular orbit, including $d_{\rm L}$, $\Omega$, and $t_{\rm c}$. 
However, full parameter estimation and uncertainty quantification via Bayesian inference is computationally expensive. 
\cite{mangiagli2020} provide parametric functions to estimate $\Delta d_{\rm L}/d_{\rm L}$, $\Delta \Omega$, and $\Delta M_{\rm tot}/M_{\rm tot}$ given the redshift, total mass, and the time to coalescence of a LISA-detected MBH merger. 
The parametric functions provided by \cite{mangiagli2020} were derived from simulating a large number of GW waveforms given some input \mtot, $z$, and $t_c$, extracting GW parameters from the waveform via Fisher information matrix evaluation, and then fitting the distributions of extracted parameters with a polynomial. 
Their simulated mergers span a mass range of $10^{5}$\,\msun\,$< M_{\rm tot} <$ $10^{7}$\,\msun~at $z < 4$. 
Other input parameters of the simulated GW waveforms used by \cite{mangiagli2020} were randomly drawn from a uniform distribution spanning a physically reasonable range. For example, the mass ratio of the MBHB was drawn from [0.1, 1] and the spin of the MBH in the binary was drawn from [0, 1]. 
We refer the reader to Section IV of \cite{mangiagli2020} for the distributions of other input GW parameters.
Using these parametric functions, we can quickly compute the localization volumes for the simulated MBH mergers in our catalogs. 

The dependence of the localization volume on the total mass and redshift of the merging system is shown in Figure~\ref{fig:errvol}, for different time points immediately before and after coalescence.
In the left panel, we keep $M_{\rm tot}$ fixed at $10^{6}$\,\msun~and allow the redshift of the merger to vary; in the middle panel, we keep the redshift fixed at $0.5$ and allow $M_{\rm tot}$ to vary. 
In both cases, the localization volume decreases as coalescence approaches, as expected. 
Furthermore, at a fixed $t_{\rm c}$, $V_{\rm loc}$ increases with both $M_{\rm tot}$ and $z$. 
The right panel shows how the size of the localization volumes depends on $M_{\rm tot}$ and $z$ of the merging system after coalescence. 
The localization volume is smallest for mergers with \mtot~$\sim10^{6.2}\,$\msun~at $z < 1$, and the \mtot~corresponding to the smallest localization volume shifts towards $\sim10^{5.8}\,$\msun~for mergers at $z > 1$.

\subsection{Contaminating Extragalactic Transients}\label{subsec:commonTrans}
\begin{figure}
    \vspace{0.1cm}
    \centering
    \includegraphics[width=.95\linewidth]{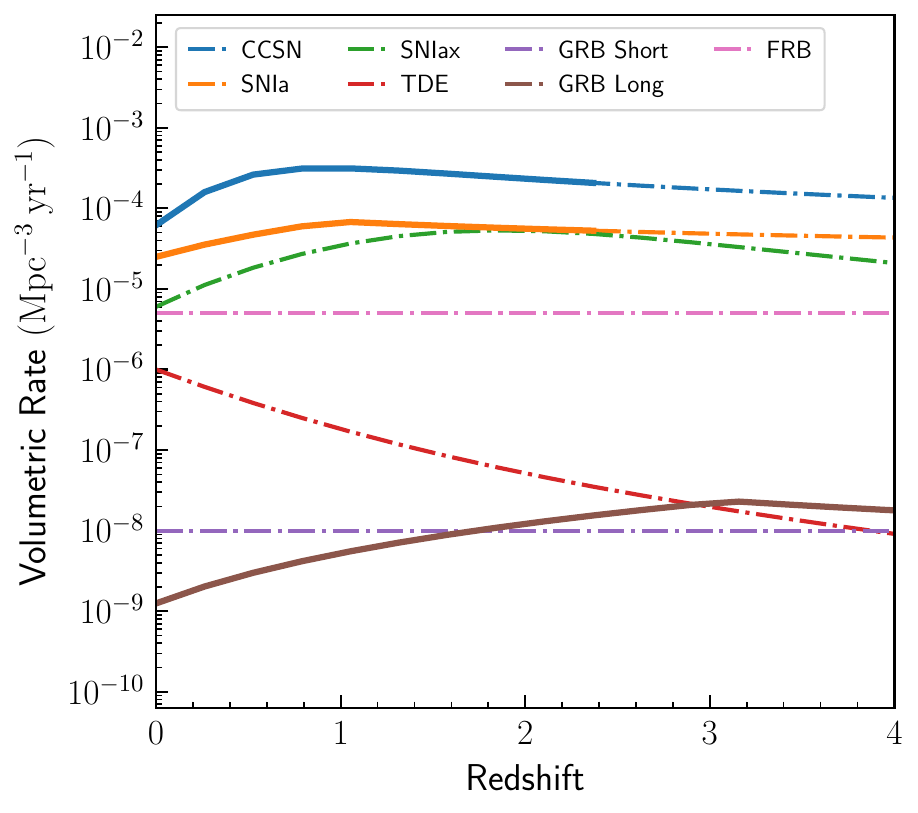}
    \caption{The adopted volumetric rate as a function of redshift for each EM transient class considered in this work. 
    Solid curves indicate the redshift range where the volumetric rate has been explicitly calibrated using observations, while dash-dotted curves indicate the redshift range where the volumetric rate is derived using indirect correlations or theoretical models.
    }
    \label{fig:transVolRate}
\end{figure}

We consider various classes of common extragalactic transients that could act as contaminants in a telescope follow-up search for the transient EM counterparts of LISA MBH mergers.
Specifically, we include type Ia supernova (SNIa), type Iax supernova (SNIax),  core-collapse supernova (CCSN), gamma-ray burst (GRB), tidal disruption event (TDE), and fast radio burst (FRB) in our list. 
Transients from these classes also have a peak luminosity that is comparable to the luminosity of actively accreting MBHs. 
For example, a $10^5\,$\msun~MBH accreting at the Eddington limit has a bolometric luminosity of $\sim10^{43}\,\rm\,erg\,s^{-1}$, which is similar to the peak luminosity of a SNIa~\citep{Phillips1993, Hamuy1996}.  
In the rest of this section, we list the volumetric rate (in \mpcyr) adopted to simulate each class of extragalactic transients. 
We also plot the volumetric rate of these common extragalactic transients and their redshift dependence in Figure~\ref{fig:transVolRate}.

Type Ia supernovae are thermonuclear explosions of carbon-oxygen white dwarf (WD) stars. Type Ia supernovae reach a peak luminosity of $\sim10^{42-43}$ $\rm erg\,s^{-1}$ in the UV, optical, and infrared~\citep{Phillips1993, Hamuy1996}. 
We use the volumetric rate from \cite{Dilday2008} to simulate SNIa events at $z < 1$, and the volumetric rate from \cite{Hounsell2018} for simulated SNIa at $z > 1$: 
\begin{equation}
         R_{\rm SNIa}(z)=
        \begin{cases}
            2.5\times10^{-5}(1+z)^{1.5}\:\rm{Mpc^{-3}\:yr^{-1}} & z < 1,\\
            9.7\times10^{-5}(1+z)^{-0.5}\:\rm{Mpc^{-3}\:yr^{-1}} & z > 1.
        \end{cases}
    \label{eqn:snla_rate}
\end{equation}

Type Iax supernovae are the largest class of peculiar WD supernovae. Type Iax supernovae share some similarities with normal Type Ia supernovae, but in general have a lower peak luminosity~\citep{Foley2013}. The volumetric rate of SNIax is set to $6\times10^{-6}\:\rm{Mpc^{-3}\:yr^{-1}}$ at $z=0$~\citep{Foley2013}, and the redshift evolution of this rate is chosen to obey the star formation rate given in \cite{Madau2014}:
\begin{equation}
    R_{\rm SNIax}(z)= 6\times10^{-6}\frac{(1+z)^{2.7}}{1+[(1+z)/2.9]^{5.6}}\:\rm{Mpc^{-3}\:yr^{-1}}.
\end{equation}

Core-collapse supernovae are the explosions of massive stars at the end of their lifetime, and they also reach a peak luminosity of $\sim10^{42-43}$ $\rm erg\,s^{-1}$~\citep{Smartt2009}. CCSN are the most abundant class of supernovae. We adopt the volumetric rate from \cite{Strolger2015}:
\begin{equation}
    R_{\rm CCSN}(z)= 1.365\times10^{-4}\frac{(1+z)^{5}}{1+[(1+z)/1.5]^{6.1}}\:\rm{Mpc^{-3}\:yr^{-1}}.
\end{equation}

Gamma-ray bursts are short and intense bursts of gamma-rays that are commonly observed in the high-energy sky~\citep{Gehrels2004}. 
GRBs might be confused with the high-energy emission from MBH mergers at/after coalescence~\citep[e.g.,][]{Yuan2021}. 
Furthermore, their afterglow emission across the EM spectrum~\citep{vanParadijs2000, Granot2002} may be confused with the post-merger emission of LISA-detected coalescing MBHBs~\citep[e.g.,][]{Lippai2008, Rossi2010, Zanotti2010}. 
For short GRBs, we adopt the observed volumetric rate from \cite{Fong2015}:
\begin{equation}
    R_{\rm GRB_{S}} = 1\times10^{-7}\:\rm{Mpc^{-3}\:yr^{-1}}.
\end{equation}
For long GRBs, we adopt the redshift-dependent volumetric rate from \cite{Wanderman2010}:
\begin{equation}
    R_{\rm GRB_{L}}(z) = R_0
    \begin{cases}
        (1+z)^{2.07} & z < z_1,\\
        (1+z_1)^{3.43}(1+z)^{-1.36} & z > z_1,
    \end{cases}    
\end{equation}
where $R_0 = 1.25\times10^{-9}\:\rm{Mpc^{-3}\:yr^{-1}}$ and $z_1 = 3.11$. 

Tidal disruption events occur when a star passes near a MBH, and the strong tidal force acting on the star tears it apart~\citep{Rees1988}. 
TDEs are less common than supernovae, but they preferentially occur at the centers of galaxies, and thus the chance of confusion could be high. 
For TDEs, the volumetric rate at $z = 0$ is taken from \cite{vanVelzen2018}, and we adopt the redshift evolution of this rate from \cite{Kochanek2016}:\footnote{The $z$-dependence of TDE's volumetric rate is derived from fitting the curves in Figure 14 of \cite{Kochanek2016} with a third order polynomial. The constant term of the best-fit polynomial is omitted and replaced by the volumetric rate at $z\simeq0$ from \cite{vanVelzen2018}.}
\begin{eqnarray}
    R_{\rm TDE}(z) &&= 1\times10^{-6}\times10^{f(z)}\:\rm{Mpc^{-3}\:yr^{-1}},\nonumber\\
    f(z) &&= -0.009\,z^3 + 0.121\,z^2 + 0.850\,z.
\end{eqnarray}

Fast radio bursts are millisecond-duration radio pulses that originate primarily from unknown extragalactic sources~\citep{Thornton2013, Petroff2019}. 
They are common in the radio sky, and their transient nature can be confused with predicted prompt emission from MBH mergers right before coalescence~\citep[e.g.,][]{Palenzuela2010, Kelly2017}. 
We adopt a volumetric rate of $R_{\rm FRB} \approx 5\times10^{-5}\:\rm{Mpc^{-3}\:yr^{-1}}$ from \cite{Law2017}, where an average beaming fraction of $0.1$ is assumed.

\begin{figure*}
    \centering
    \epsscale{1.16}
    \plotone{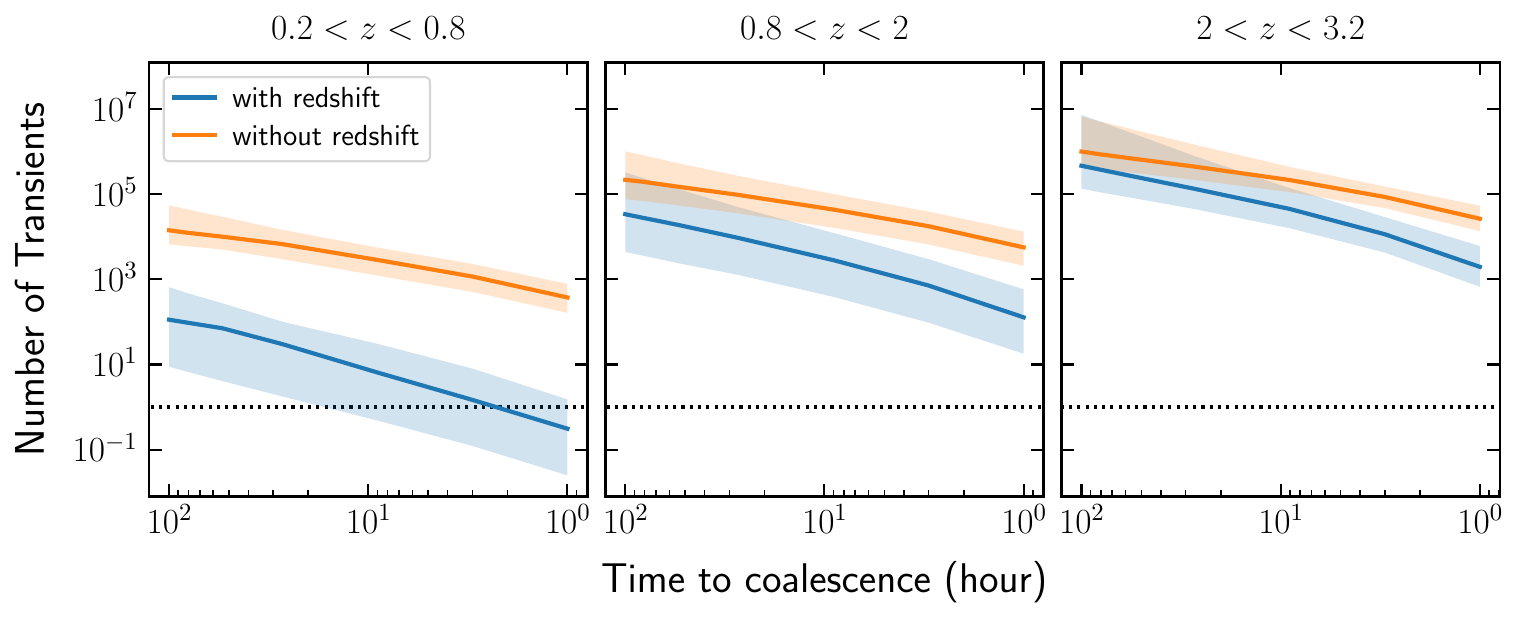}
    \caption{
    Results of the first investigation---the average number of unrelated transients expected in LISA localization volumes at different times before coalescence. 
    The averaging is taken over simulated MBH mergers in our first catalog that fall into the redshift ranges indicated at the top of each panel. 
    The blue curves give the estimated number of unrelated transients assuming redshifts are available for all transients detected in LISA sky localization regions, so that it is possible to exclude transients that reside outside LISA localization volumes using their redshifts.
    In contrast, the orange curves give the estimated number of unrelated transients assuming redshifts are not available for transients detected in the same sky region.
    The shaded regions indicate the 1-$\sigma$ range{~(central 68\%)} of the distribution for the number of unrelated transients.
    The horizontal dotted lines mark a single transient, for reference.
    Redshift information is critical to reduce the average number of unrelated (potentially contaminating) transients.
    The average number of unrelated transients in LISA localization volumes only drops down to unity for low-redshift MBH mergers (left panel) and at a few hours before coalescence.
    }
    \label{fig:nTrans_avg}
\end{figure*}

\subsection{Expected Number of Unrelated EM Transients Coinciding with each LISA MBH Merger Event}\label{subsec:totalRates}

We first consider an idealized scenario where it is possible to obtain redshifts for all EM transients detected in the LISA sky localization regions.
In this case, the expected number of unrelated EM transients that coincide with a LISA MBH merger event is the product of three quantities: the size of the merger's localization volume, the volumetric rate(s) of unrelated transients, and the time period over which these transients are observable. 

Since the redshift spanned by a typical LISA localization volume is often non-negligible and the volumetric rates of unrelated EM transients considered here have redshift dependence, we integrate the product of the volumetric rate $R(z)$, the differential comoving volume $dV_{\rm c}(z)$, and a rest-frame time period $\Delta t$, over the redshift range extended by the localization volume to calculate the expected number of unrelated transients:
\begin{equation}\label{eqn:Ntrans}
    N_{\rm trans} = \frac{\Delta \Omega}{4\pi}\int_{z_{\min}}^{z_{\max}} R(z)\,{dV_{\rm c}(z)}\,{\Delta t}\,dz.
\end{equation}
Similar to Equation~\ref{eqn:vol}, $z_{\min}$ is the redshift at the luminosity distance of $d_{\rm L} - \Delta d_{\rm L}$ and $z_{\max}$ is the redshift at $d_{\rm L} + \Delta d_{\rm L}$.
{Note that $\Delta d_{\rm L}$ and $\Delta \Omega$ are functions of $t_{\rm c}$, thus, Equation~\ref{eqn:Ntrans} computes the expected number of unrelated transients at a given $t_{\rm c}$.}

\begin{figure*}
    \epsscale{1.16}
    \plotone{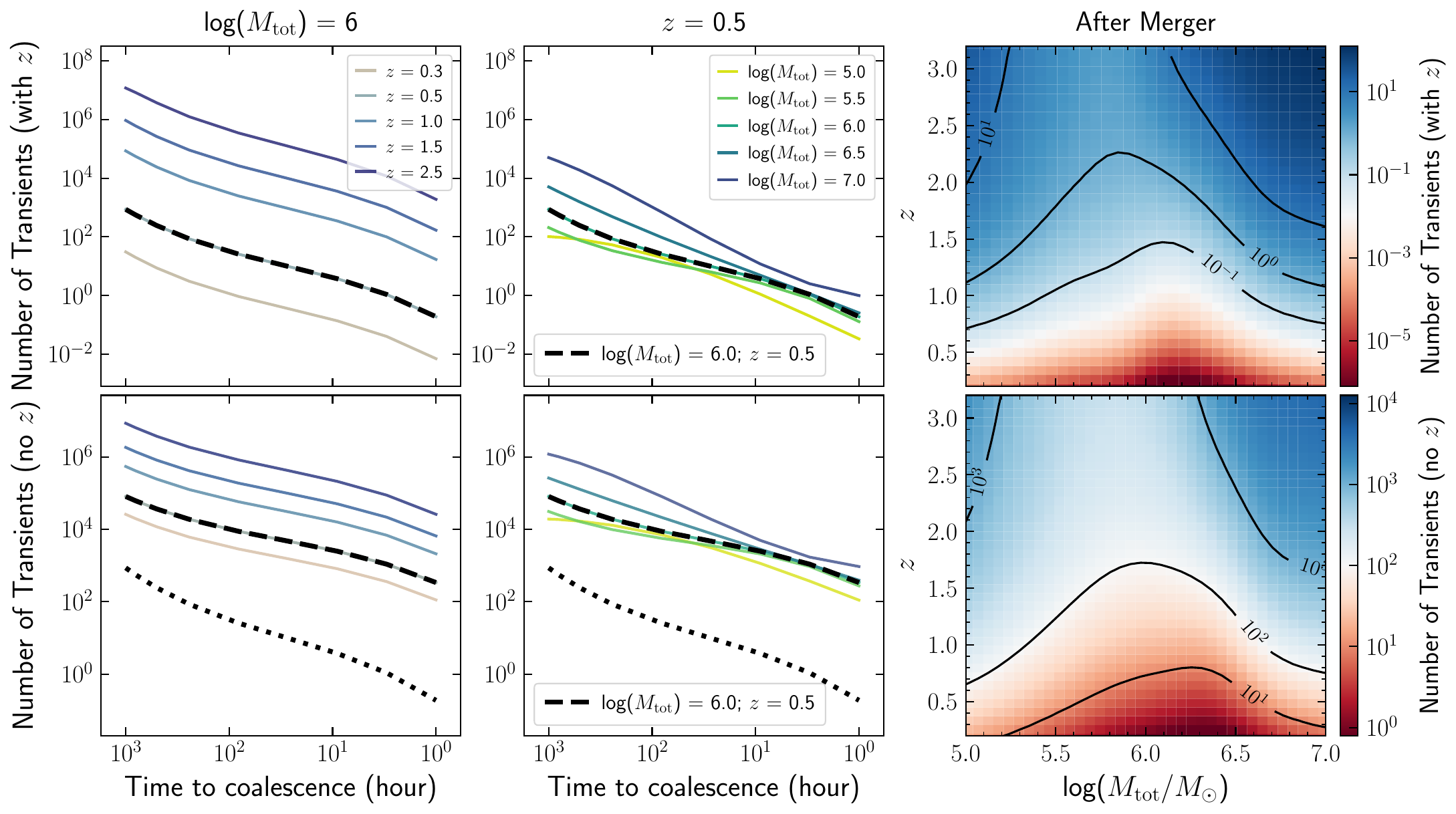}
    \caption{
    The number of unrelated transients coinciding with LISA MBH merger events, as a function of total binary mass (\mtot) and redshift ($z$), at different times before and after coalescence.
    {\it Top}: The left and center panels show the number of unrelated transients in LISA localization volumes, at different times before coalescence.
    Here, we assume that redshifts for all transients detected in LISA sky localization regions are available, such that transients residing outside LISA localization volumes can be excluded. 
    In the left panel, \mtot~is fixed at $10^6\,$\msun; in the center panel, the redshift is fixed at $z = 0.5$.
    The dashed lines show the number of unrelated transients for a reference MBH merger with $M_{\rm tot} = 10^{6}$~\msun~at $z=0.5$.
    The right panel shows the corresponding numbers after coalescence.
    The trends of the expected number of unrelated transients with \mtot, $z$, and $t_c$ are similar to that of the localization volumes as shown in Figure~\ref{fig:errvol}.    
    {\it Bottom}: Same as the plots in the top row, but assuming redshifts are not available, thus every unrelated transient detected in LISA sky localization regions from $z = 0$ to $3$ is included~(see Section~\ref{subsec:totalRates}).
    To help guide the eye, the dotted lines in the left and center panels show the number of unrelated transients assuming redshifts are available for all detected transients, for a reference merger with $M_{\rm tot} = 10^{6}$~\msun~at $z=0.5$.
    A comparison of the dashed and dotted lines shows that the number of unrelated transients will increases by a factor of $\sim100$ without redshifts for the detected transients.
    }
    \label{fig:results_trends}
\end{figure*}

Given the observing strategy we assumed in Section~\ref{sec:bkg}, $\Delta t$ is equivalent to the characteristic lifetime of the transient.
We take the median transient lifetime of SNIa, SNIax, CCSN, and GRBs from \cite{Villar2017}. 
The lifetimes adopted for short and long GRBs are much longer than their duration in gamma-rays, since we are taking into account the duration of their synchrotron afterglows.
For TDEs, we adopt the median duration of the TDEs included in \cite{Yao2023}.
For FRBs, since its typical duration of $\sim3$ milliseconds~\citep{Petroff2019} is much shorter than the integration time of most general-purpose radio transient surveys~\citep[e.g.,][]{Murphy2013, Lacy2020}, we set $\Delta t = T_{\rm obs}/(1+z)$ with $T_{\rm obs}$ being the integration/observation time.
The number of expected FRBs simply scales linearly with the adopted $T_{\rm obs}$.
Without loss of generality, we adopt an integration/observation time of 3 minutes.
In Section~\ref{subsec:result_class}, we will show the effect of increasing the integration time to 1 hour. 
We summarize the adopted $\Delta t$ for our considered transients (aside from FRBs) in Table~\ref{tab:transLF}. 

We next consider a more realistic scenario in which the redshifts for most EM transients detected in the LISA sky localization regions are not available.
In this case, we must count all extragalactic transients that are detectable in the sky localization region ($\Delta \Omega$) of the LISA MBH merger event. 
Thus, to compute the number of expected contaminating transients, we perform an integration of Equation~\ref{eqn:Ntrans} from $z_{\min} = 0$ to $z_{\max} = 3$. 
Note that we choose this high redshift cut to exclude common transients that are too faint to be detectable even with the largest telescopes.

\begin{table}
    \vspace{0.2cm}
    \centering\caption{Typical Lifetime of Common Transients\label{tab:transLF}}
    \vspace{-0.2cm}
    \centering{
    \begin{tabular}{lcc}
    \hline\hline
    Transient Class & Lifetime, $\Delta t$ & Reference\\
                    & [day] & \\
    \hline
    SN Type Ia & 25 & \cite{Villar2017}\\
    SN Type Iax & 35 & \cite{Villar2017}\\
    Core-collapse SN & 55 & \cite{Villar2017}\\
    Short GRBs & 1 & \cite{Villar2017}\\
    Long GRBs & 20 & \cite{Villar2017}\\
    TDEs & 56 & \cite{Yao2023}\\
    \hline\hline
    \end{tabular}
    }
\end{table}

\section{Results} \label{sec:results}

\begin{figure*}
    \centering
    \epsscale{1.16}
    \plotone{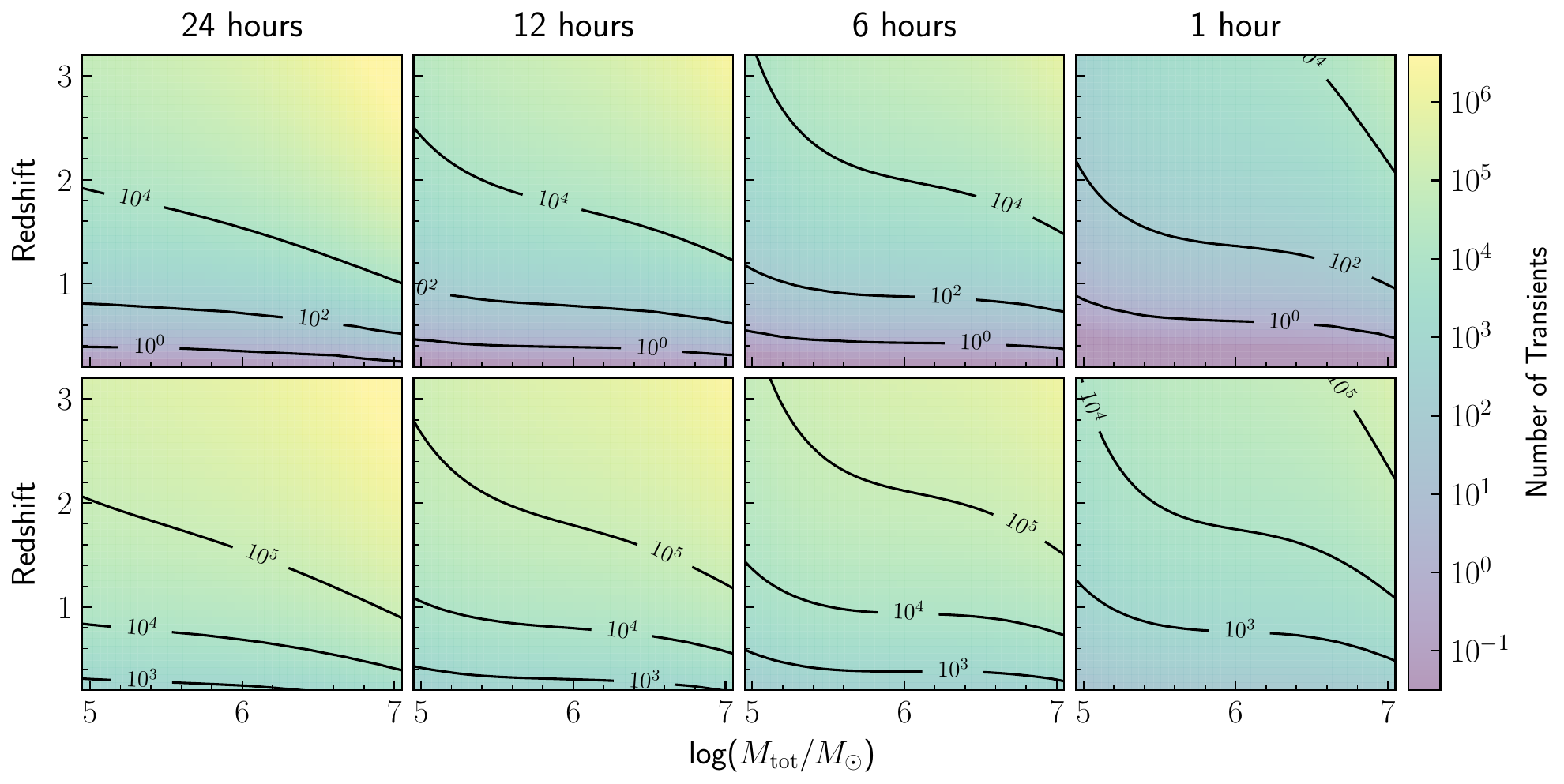}
    \caption{
    The number of unrelated transients coinciding with LISA MBH merger events, as a function of total binary mass (\mtot) and redshift ($z$), at different times before coalescence.    
    The colors and contour lines indicate the expected number of unrelated transients for MBH mergers with a particular combination of $M_{\rm tot}$ and $z$. 
    {\it Top:} The number of unrelated transients assuming redshifts are available for all transients detected in LISA sky localization regions, where redshifts are used to exclude transients that reside outside LISA localization volumes.
    The number of unrelated transients drops to unity for MBH mergers at $z\lesssim0.2$ at more than 6 hours before coalescence, and drops to unity for MBH mergers at $z\lesssim0.7$ at 1 hour before coalescence.
    {\it Bottom:} The number of unrelated transients assuming redshifts are not available for the transients detected in LISA sky localization regions.
    Without redshifts, the number of unrelated transients increases significantly for all combinations of \mtot, $z$, and $t_c$, and it does not drop below unity even for the lowest redshift mergers.
    }
    \label{fig:results_map}
\end{figure*}

\subsection{Investigation \#1: The Average Number of Unrelated Transients in LISA Localization Volumes}\label{subsec:result_avg}

We compute the average number of unrelated transients in LISA localization volumes, and find that redshift information for detected transients is critical for reducing the number of potential contaminants down to unity.
Figure~\ref{fig:nTrans_avg} shows the estimated number of unrelated transients for simulated MBH mergers in three different redshift ranges. The redshift ranges are chosen for illustration purposes, and the dependence of the number of unrelated transients on MBH merger redshift is further examined in Section~\ref{subsec:result_param_space}.
{For each simulated MBH merger, the expected number of unrelated transients is evaluated at a range of $t_{\rm c}$ (from 1000 hours to 1 hour).}
The solid line presents the average number of unrelated transients associated with simulated MBH mergers in the corresponding redshift range, and the shaded region indicates the 1-$\sigma$ range {(central 68\%)} for the distribution.
The blue curves show the numbers for the idealized scenario where redshifts are available for all detected transients; in contrast, the orange curves show the numbers for the scenario where redshifts are not available for the detected transients.
With redshift information, the average number of potential contaminants only drops to unity for low-redshift (the left panel) MBH mergers, and at a few hours before coalescence.
Without redshift information, the number of potential contaminants increase by a factor of $\sim$~100 on average, and it never drops below unity.

\subsection{Investigation \#2: A Parameter Space Study of Contaminating Transients in LISA Localization Volumes}\label{subsec:result_param_space}

In this section, we perform our second investigation, which examines how the expected number of unrelated transients in LISA localization volumes change as a function of $M_{\rm tot}$ and $z$ of the detected MBH mergers.
This second study extends our first investigation in Section~\ref{subsec:result_avg}, to identify the parameter space of $M_{\rm tot}$~and~$z$ where the number of unrelated and potentially contaminating transients is the lowest. 

The number of unrelated transients in the idealized scenario where redshift information is available for all detected transients drops to unity for relatively low-redshift MBH mergers as early as hours before coalescence.
The left two panels in the top row of Figure~\ref{fig:results_trends} show the expected number of transients as a function of the time to coalescence.
We keep \mtot~fixed at $10^6\,$\msun~and allow the redshift to vary in the top-left panel, and we keep the redshift fixed at $z = 0.5$ and allow \mtot~to vary in the top-middle panel. 
These two panels show that, in the pre-merger phase, the number of contaminating transients drops to unity for low-redshift mergers as early as hours before coalescence. 
We further explore how the number of transients change with $M_{\rm tot}$ and $z$ at 24, 12, 6, and 1 hour before coalescence, and the results are shown by the top row of Figure~\ref{fig:results_map}.
At more than 6 hours before coalescence, the number of unrelated transients drops to unity for MBH mergers of all masses at $z\lesssim0.2$. 
At the final hour before coalescence, the number of unrelated transients drops to unity for MBH mergers of all masses at $z\lesssim0.6$.
After coalescence, the expected number of transients is shown in the top-right panel of Figure~\ref{fig:results_trends}. 
The number of unrelated transients is below unity for MBH mergers with $10^{5.5}$\,\msun~$< M_{\rm tot} <$~$10^{6.5}\,$\msun~at $z \lesssim 1.5$.
{Overall, the number of unrelated transients follows similar trends with \mtot~and~$z$ as that of LISA GW localization volume (see Figure~\ref{fig:errvol}), which suggests that the intrinsic number of unrelated transients in LISA localization volume is primarily determined by the size of the localization volume, while the $z$-dependence of unrelated transients' volumetric rates has a negligible impact.
}

In contrast, in the scenario where redshift information is not available for the transients detected in LISA sky localization regions, the expected number of unrelated transients increases by a factor of $\sim100$.
The bottom row of Figure~\ref{fig:results_trends} shows the results for this scenario, where we count all transients detected in the LISA sky localization region from $z_{\min} = 0$ to $z_{\max} = 3$.
The dashed curves in the left and center panels highlight the number of unrelated transients for a MBH merger with $M_{\rm tot} = 10^{6}$~\msun~at $z=0.5$. 
As reference, the dotted curves show the corresponding number of unrelated transients for the same merger, assuming the redshift information for all detected transients is available.
Thus, the difference between the dashed and the dotted curves illustrates that the number of unrelated and potentially contaminating transients increases by an average factor of $\sim100$ without redshift information.
In the bottom row of Figure~\ref{fig:results_map}, we show the number of transients as a function of $M_{\rm tot}$ and $z$ at 24, 12, 6, and 1 hour before coalescence.
By comparing the panels in the bottom row of Figure~\ref{fig:results_map} with the panels in the top row of Figure~\ref{fig:results_map}, it is again shown that the number of unrelated transients increases significantly without redshift information.
This conclusion remains for the expected number of transients after coalescence by comparing the bottom-right and top-right panels of Figure~\ref{fig:results_trends}.

\subsection{Number of Unrelated Transients \\vs. Transient Class}\label{subsec:result_class}
\begin{figure}
    \vspace{0.2cm}
    \centering
    \epsscale{1.1}
    \plotone{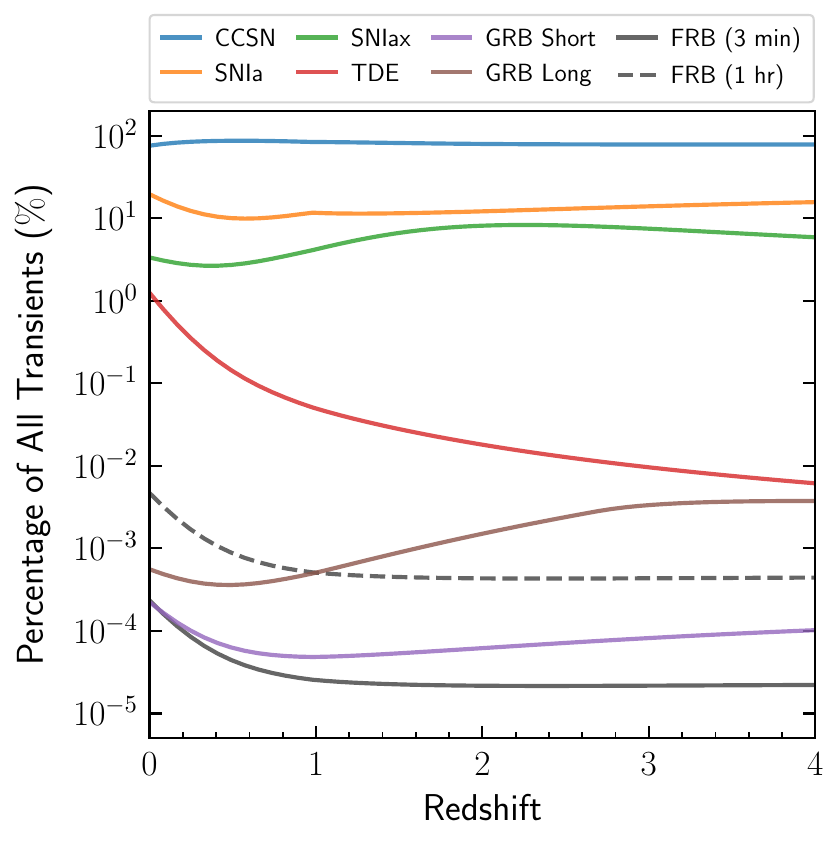}
    \caption{The percentage of contaminating transients belonging to a particular class as a function of redshift. Core-collapse supernovae account for most of the contaminating transients.  
    The numbers of expected GRBs and FRBs are negligible in comparison to that of core-collapse supernovae.
    The solid gray line shows the percentage of FRBs for an integration/observation time of 3 minutes, and dashed gray line shows the percentage of FRBs for an integration/observation time of 1 hour. 
    }
    \label{fig:results_transClass}
\end{figure}

Lastly, we consider the percentage of unrelated transients belonging to different transient classes, and find that the entire population is dominated by supernovae, particularly CCSN.
Figure~\ref{fig:results_transClass} shows that CCSN dominate over all other transient classes at all $z$, and account for $\sim80\%$ of all unrelated transients considered.
SNIa and SNIax rank second and third, and account for $\sim13\%$ and $\sim6\%$ of the unrelated transients, respectively.
{TDEs account for $\sim1\%$ of all unrelated transients in the local Universe, and its share drops below $\sim0.01\%$ beyond $z\sim3$.
The shares of short GRBs and FRBs are negligible ($<0.0001\%$), and the share of long GRBs is $<0.003\%$ over the redshift range considered.}
The gray dashed line in Figure~\ref{fig:results_transClass} shows the percentage of FRBs assuming an integration/observation time ($T_{\rm obs}$) of 1 hour, which is still negligible in comparison to the dominant classes of unrelated transients.

\section{Discussion} \label{sec:discuss}
\subsection{Associating EM Transients with MBH Mergers Detected By LISA}

Redshift information for detected transients is vital to successful and timely identification of the true transient EM counterparts to MBH mergers detected by LISA.
Our results presented in Section~\ref{sec:results} demonstrate that the number of potential contaminants increases by an average factor of $\sim100$ without adequate redshift information for transients detected in LISA sky localization regions. 
For example, the number of potential contaminants is $\sim1000$ for a MBH merger at $z \sim0.8$ and at 1 hour before coalescence (bottom-right panel of Figure~\ref{fig:results_map}), assuming redshifts for the detected transients are not available.
In contrast, the number of potential contaminants will drop to $\sim3$ for the same MBH merger if redshifts are available for all detected transients (top-right panel of Figure~\ref{fig:results_map}).
Therefore, without redshifts for the detected transients, locating the true transient EM counterparts of LISA MBH merger events will be challenging. 
Upcoming (and ongoing) wide-field multi-band imaging and spectroscopic surveys will create deep and detailed maps of our Universe up to high redshifts~\citep[e.g.,][]{WFIRST2015, desicollaboration2016, Lou2016, Euclid2017, Zhan2018, ivezic2019}.
Reliable photometric and spectroscopic redshifts~\citep[e.g.,][]{Loh1986, Connolly1995, Bolzonella2000, CarrascoKind2013} obtained for galaxies covered by these surveys will help provide redshifts for transients detected in LISA localization regions, assuming detected transients can be robustly associated with their host galaxies~\citep[e.g.,][]{Gagliano2021, Gagliano2023}. 
In real-time observations, light curve modeling can also provide redshifts for the detected transients~\citep{Sullivan2006, sako2008, Kessler2010, Qu2023}. 
Given their redshift information, unrelated transients residing outside LISA localization volumes can be easily excluded, and the number of potential contaminants will be greatly reduced.



Robust transient classification algorithms/pipelines could help rule out potential contaminants.
Our dynamic sky is monitored by many time-domain surveys covering a wide range of the EM spectrum~\citep[e.g.,][]{Atwood2009, ivezic2019, CHIME/FRBCollaboration2018, Lacy2020, Predehl2021}. 
When these surveys detect a change in the brightness at any location on the sky, they will send out alerts to the community and those alerts will be collected and combined at the alert brokers~\citep[e.g.,][]{Narayan2018, Forster2021, Moller2021, Matheson2021}. 
The light curve and image stamps included in the alert package are then used to classify the associated source into different classes of known physical natures~\citep[e.g.,][]{Gomez2020, Carrasco-Davis2021, Forster2021} or anomalies~\citep[e.g.,][]{Villar2021, Martinez-Galarza2021, Muthukrishna2021, Perez-Carrasco2023}, using algorithms deployed onto the alert brokers.
If a large fraction of transients detected in LISA localization regions can be confidently classified into classes of known natures (e.g., SNIa), the number of potential contaminants can also drop substantially.

\subsection{Caveats \& Future Work}

The simulated MBH mergers included in our catalogs have only two free parameters (\mtot~and $z$), thus future work should attempt to investigate how our results depend on other GW parameters (e.g., mass ratio).
GW signals of MBH mergers detected by LISA are commonly defined by a total of 15 parameters~\citep{Amaro-Seoane2017, colpi2024}.
The reason we only allow \mtot~and $z$ to vary for our simulated MBH mergers is that the method we adopt to quickly compute LISA localization volumes only takes these two inputs~\citep{mangiagli2020}, and adopts a uniform distribution for the other parameters. 
It will be worthwhile to rerun the experiment carried out by \cite{mangiagli2020} with added dependencies on additional GW parameters, and explore the effect of these parameters on the predicted numbers of contaminating transients.

Real telescope follow-up of LISA events will be further complicated by many additional factors, such as the detectability of a contaminating transient given the flux limit of follow-up observations and the actual observing cadence.
For example, a $z\sim2$ Type Ia supernova can be detected by LSST but not the Zwicky Transient Facility~\citep[ZTF;][]{ZTF2019a, ZTF2019b} given the shallow flux limit of ZTF. 
Nonetheless, the high observing cadence of ZTF allows it to discover many nearby and luminous supernovae that might saturate the detectors of LSST.
Furthermore, if the detected transient coincides with the nucleus of a galaxy, TDEs are the most likely contaminants, thus the total expected number of unrelated transients will drop significantly. 
Future investigation should fold in the actual observing scenarios in order to provide the most robust estimates for the expected number of contaminating transients.

{More realistic merger rate predictions, as a function of \mtot~and~$z$, will help better characterize the average number of unrelated EM transients in LISA localization volumes. 
In Section~\ref{subsec:result_avg}, we adopt the merger rate distribution from the NewHorizon simulation, where the MBHs are seeded with a fixed mass of $10^4\,$\msun~\citep{Volonteri2020}. 
Howeber, MBH seed masses could have very different distributions depending on the assumed MBH formation mechanisms~(see \citealt{Volonteri2010} and references therein).
Furthermore, the NewHorizon simulation does not have the resolution to fully simulate all of the physics involved. For instance, the effect of dynamical friction on binary evolution is added in post-processing.
Merger rate predictions from future cosmological simulations that can self-consistently incorporate all of the relevant physics and a range of MBH formation mechanisms will help better estimate the average number of contaminating transients in LISA localization volumes.}


\section{Conclusions}\label{sec:conclude}
In this work, we perform calculations to estimate the number of unrelated EM transients that are expected to occur in the GW localization volumes of MBH mergers detected by LISA.
Those unrelated transients will contaminate telescope follow-up searches for transient EM counterparts of LISA MBH mergers.
Quantifying the number of unrelated transients is a first step toward developing effective strategies for telescope follow-up that aim to identify and observe the true transient EM counterparts. Our main results are summarized below:

\begin{itemize}
    \item Before coalescence, the localization volume of MBH mergers detected by LISA increases with the total mass and redshift of the mergers, and decreases with the time to coalescence. 
    After coalescence, the MBH mergers with \mtot$\,\approx10^{6}\,$\msun~have the smallest localization volumes.

    \item Redshift information for all transients detected in the LISA sky localization region is required to bring the total number of contaminating transients down to unity. The number of contaminating transients increases by an average factor of $\sim100$ in the case of no redshift information in comparison to the case where the redshift of every detected transient is available. 

    \item In the most idealized case where the redshifts of all detected transients are available, the number of contaminating transients drops to unity as soon as 6 hours before coalescence for mergers at $z \lesssim 0.2$. 
    After coalescence, the number of unrelated transients is below unity for mergers with $10^{5.5}$\,\msun~$< M_{\rm tot} <$~$10^{6.5}$\msun~at $z \lesssim 1.5$.
    \end{itemize}

The work presented here represents a first attempt to assess the challenge of identifying transient EM counterparts to MBH mergers detected by LISA. 
Endeavors focusing on developing telescope follow-up strategies for LISA GW events should focus on three areas: (1) building large redshift catalogs for host galaxies, (2) developing robust real-time transient classification algorithms that will precisely identify transients of known natures, (3) coordinating telescope resources that will allow obtaining redshifts for a larger number of transients in a timely manner. 
We also note that the investigations carried out in this work only probes two extreme scenarios in redshifts. The planned future work based on more realistic simulations of follow-up observations will provide a more accurate estimate for the expected number of contaminating transients.

\begin{acknowledgments}
We thank the anonymous referee for their thoughtful comments that helped us improve the paper. W.Y. acknowledges support from the Dunlap Institute for Astronomy \& Astrophysics at the University of Toronto. J.J.R.\ and D.H.\ acknowledge support from the Canada Research Chairs (CRC) program, the NSERC Discovery Grant program, the NSERC Alliance International program, the Canada Foundation for Innovation (CFI). J.J.R.\ also acknowledges the Qu\'{e}bec Minist\`{e}re de l’\'{E}conomie et de l’Innovation.
J.C.R and C.D. acknowledge funding from the National Aeronautics and Space Administration (NASA) under award No. 80NSSC24K0439.
C.D. also acknowledges support from the NASA FINESST grant No. 80NSSC24K1478.
M.E., K.S., T.B., and K.F. acknowledge support from NASA under the NASA LISA Preparatory Science Program grant 80NSSC22K0748.
\end{acknowledgments}


\software{
        pandas \citep{pandas2010},
        numpy \citep{numpy2020},    
        scipy \citep{scipy2020},
        matplotlib \citep{matplotlib2007},
        astropy \citep{astropy2013, astropycollaboration2022},  
}

\bibliography{My_Library, LISA-I}{}

\begin{thebibliography}{}
\expandafter\ifx\csname natexlab\endcsname\relax\def\natexlab#1{#1}\fi
\providecommand{\url}[1]{\href{#1}{#1}}
\providecommand{\dodoi}[1]{doi:~\href{http://doi.org/#1}{\nolinkurl{#1}}}
\providecommand{\doeprint}[1]{\href{http://ascl.net/#1}{\nolinkurl{http://ascl.net/#1}}}
\providecommand{\doarXiv}[1]{\href{https://arxiv.org/abs/#1}{\nolinkurl{https://arxiv.org/abs/#1}}}

\bibitem[{{Abbott} {et~al.}(2017){Abbott}, {Abbott}, {Abbott}, {Acernese}, {Ackley}, {Adams}, {Adams}, {Addesso}, {Adhikari}, {Adya}, \& et~al.}]{Abbott2017}
{Abbott}, B.~P., {Abbott}, R., {Abbott}, T.~D., {et~al.} 2017, \nat, 551, 85, \dodoi{10.1038/nature24471}

\bibitem[{{Amaro-Seoane} {et~al.}(2017){Amaro-Seoane}, {Audley}, {Babak}, {Baker}, {Barausse}, {Bender}, {Berti}, {Binetruy}, {Born}, {Bortoluzzi}, {Camp}, {Caprini}, {Cardoso}, {Colpi}, {Conklin}, {Cornish}, {Cutler}, {Danzmann}, {Dolesi}, {Ferraioli}, {Ferroni}, {Fitzsimons}, {Gair}, {Gesa Bote}, {Giardini}, {Gibert}, {Grimani}, {Halloin}, {Heinzel}, {Hertog}, {Hewitson}, {Holley-Bockelmann}, {Hollington}, {Hueller}, {Inchauspe}, {Jetzer}, {Karnesis}, {Killow}, {Klein}, {Klipstein}, {Korsakova}, {Larson}, {Livas}, {Lloro}, {Man}, {Mance}, {Martino}, {Mateos}, {McKenzie}, {McWilliams}, {Miller}, {Mueller}, {Nardini}, {Nelemans}, {Nofrarias}, {Petiteau}, {Pivato}, {Plagnol}, {Porter}, {Reiche}, {Robertson}, {Robertson}, {Rossi}, {Russano}, {Schutz}, {Sesana}, {Shoemaker}, {Slutsky}, {Sopuerta}, {Sumner}, {Tamanini}, {Thorpe}, {Troebs}, {Vallisneri}, {Vecchio}, {Vetrugno}, {Vitale}, {Volonteri}, {Wanner}, {Ward}, {Wass}, {Weber}, {Ziemer}, \& {Zweifel}}]{Amaro-Seoane2017}
{Amaro-Seoane}, P., {Audley}, H., {Babak}, S., {et~al.} 2017, arXiv e-prints, arXiv:1702.00786, \dodoi{10.48550/arXiv.1702.00786}

\bibitem[{{Amaro-Seoane} {et~al.}(2023){Amaro-Seoane}, {Andrews}, {Arca Sedda}, {Askar}, {Baghi}, {Balasov}, {Bartos}, {Bavera}, {Bellovary}, {Berry}, {Berti}, {Bianchi}, {Blecha}, {Blondin}, {Bogdanovi{\'c}}, {Boissier}, {Bonetti}, {Bonoli}, {Bortolas}, {Breivik}, {Capelo}, {Caramete}, {Cattorini}, {Charisi}, {Chaty}, {Chen}, {Chru{\'s}li{\'n}ska}, {Chua}, {Church}, {Colpi}, {D'Orazio}, {Danielski}, {Davies}, {Dayal}, {De Rosa}, {Derdzinski}, {Destounis}, {Dotti}, {Du{\c{t}}an}, {Dvorkin}, {Fabj}, {Foglizzo}, {Ford}, {Fouvry}, {Franchini}, {Fragos}, {Fryer}, {Gaspari}, {Gerosa}, {Graziani}, {Groot}, {Habouzit}, {Haggard}, {Haiman}, {Han}, {Istrate}, {Johansson}, {Khan}, {Kimpson}, {Kokkotas}, {Kong}, {Korol}, {Kremer}, {Kupfer}, {Lamberts}, {Larson}, {Lau}, {Liu}, {Lloyd-Ronning}, {Lodato}, {Lupi}, {Ma}, {Maccarone}, {Mandel}, {Mangiagli}, {Mapelli}, {Mathis}, {Mayer}, {McGee}, {McKernan}, {Miller}, {Mota}, {Mumpower}, {Nasim}, {Nelemans}, {Noble}, {Pacucci}, {Panessa}, {Paschalidis}, {Pfister}, {Porquet},
  {Quenby}, {Ricarte}, {R{\"o}pke}, {Regan}, {Rosswog}, {Ruiter}, {Ruiz}, {Runnoe}, {Schneider}, {Schnittman}, {Secunda}, {Sesana}, {Seto}, {Shao}, {Shapiro}, {Sopuerta}, {Stone}, {Suvorov}, {Tamanini}, {Tamfal}, {Tauris}, {Temmink}, {Tomsick}, {Toonen}, {Torres-Orjuela}, {Toscani}, {Tsokaros}, {Unal}, {V{\'a}zquez-Aceves}, {Valiante}, {van Putten}, {van Roestel}, {Vignali}, {Volonteri}, {Wu}, {Younsi}, {Yu}, {Zane}, {Zwick}, {Antonini}, {Baibhav}, {Barausse}, {Bonilla Rivera}, {Branchesi}, {Branduardi-Raymont}, {Burdge}, {Chakraborty}, {Cuadra}, {Dage}, {Davis}, {de Mink}, {Decarli}, {Doneva}, {Escoffier}, {Gandhi}, {Haardt}, {Lousto}, {Nissanke}, {Nordhaus}, {O'Shaughnessy}, {Portegies Zwart}, {Pound}, {Schussler}, {Sergijenko}, {Spallicci}, {Vernieri}, \& {Vigna-G{\'o}mez}}]{Amaro-Seoane2023}
{Amaro-Seoane}, P., {Andrews}, J., {Arca Sedda}, M., {et~al.} 2023, Living Reviews in Relativity, 26, 2, \dodoi{10.1007/s41114-022-00041-y}

\bibitem[{{Armitage} \& {Natarajan}(2002)}]{Armitage2002}
{Armitage}, P.~J., \& {Natarajan}, P. 2002, \apjl, 567, L9, \dodoi{10.1086/339770}

\bibitem[{{Astropy Collaboration} {et~al.}(2022){Astropy Collaboration}, {Price-Whelan}, Lim, Earl, Starkman, Bradley, Shupe, Patil, Corrales, Brasseur, N{\"o}the, Donath, Tollerud, Morris, Ginsburg, Vaher, Weaver, Tocknell, Jamieson, {van Kerkwijk}, Robitaille, Merry, Bachetti, G{\"u}nther, Aldcroft, {Alvarado-Montes}, Archibald, B{\'o}di, Bapat, Barentsen, Baz{\'a}n, Biswas, Boquien, Burke, Cara, Cara, Conroy, Conseil, Craig, Cross, Cruz, D'Eugenio, Dencheva, Devillepoix, Dietrich, Eigenbrot, Erben, Ferreira, {Foreman-Mackey}, Fox, Freij, Garg, Geda, Glattly, Gondhalekar, Gordon, Grant, Greenfield, Groener, Guest, Gurovich, Handberg, Hart, {Hatfield-Dodds}, Homeier, Hosseinzadeh, Jenness, Jones, Joseph, Kalmbach, Karamehmetoglu, Ka{\l}uszy{\'n}ski, Kelley, Kern, Kerzendorf, Koch, Kulumani, Lee, Ly, Ma, MacBride, Maljaars, Muna, Murphy, Norman, O'Steen, Oman, Pacifici, Pascual, {Pascual-Granado}, Patil, Perren, Pickering, Rastogi, Roulston, Ryan, Rykoff, Sabater, Sakurikar, Salgado, Sanghi, Saunders,
  Savchenko, Schwardt, {Seifert-Eckert}, Shih, Jain, Shukla, Sick, Simpson, Singanamalla, Singer, Singhal, Sinha, Sip{\H o}cz, Spitler, Stansby, Streicher, {\v S}umak, Swinbank, Taranu, Tewary, Tremblay, {de Val-Borro}, Van~Kooten, Vasovi{\'c}, Verma, {de Miranda Cardoso}, Williams, Wilson, Winkel, {Wood-Vasey}, Xue, Yoachim, Zhang, Zonca, \& {Astropy Project Contributors}}]{astropycollaboration2022}
{Astropy Collaboration}, {Price-Whelan}, A.~M., Lim, P.~L., {et~al.} 2022, ApJ, 935, 167, \dodoi{10.3847/1538-4357/ac7c74}

\bibitem[{{Atwood} {et~al.}(2009){Atwood}, {Abdo}, {Ackermann}, {Althouse}, {Anderson}, {Axelsson}, {Baldini}, {Ballet}, {Band}, {Barbiellini}, \& et~al.}]{Atwood2009}
{Atwood}, W.~B., {Abdo}, A.~A., {Ackermann}, M., {et~al.} 2009, \apj, 697, 1071, \dodoi{10.1088/0004-637X/697/2/1071}

\bibitem[{{Begelman} {et~al.}(1980){Begelman}, {Blandford}, \& {Rees}}]{Begelman1980}
{Begelman}, M.~C., {Blandford}, R.~D., \& {Rees}, M.~J. 1980, \nat, 287, 307, \dodoi{10.1038/287307a0}

\bibitem[{{Behroozi} {et~al.}(2019){Behroozi}, {Wechsler}, {Hearin}, \& {Conroy}}]{Behroozi2019}
{Behroozi}, P., {Wechsler}, R.~H., {Hearin}, A.~P., \& {Conroy}, C. 2019, \mnras, 488, 3143, \dodoi{10.1093/mnras/stz1182}

\bibitem[{Bellm {et~al.}(2019)Bellm, Kulkarni, Graham, Dekany, Smith, Riddle, Masci, Helou, Prince, Adams, Barbarino, Barlow, Bauer, Beck, Belicki, Biswas, Blagorodnova, Bodewits, Bolin, Brinnel, Brooke, Bue, Bulla, Burruss, Cenko, Chang, Connolly, Coughlin, Cromer, Cunningham, De, Delacroix, Desai, Duev, Eadie, Farnham, Feeney, Feindt, Flynn, Franckowiak, Frederick, Fremling, {Gal-Yam}, Gezari, Giomi, Goldstein, Golkhou, Goobar, Groom, Hacopians, Hale, Henning, Ho, Hover, Howell, Hung, Huppenkothen, Imel, Ip, Ivezi{\'c}, Jackson, Jones, Juric, Kasliwal, Kaspi, Kaye, Kelley, Kowalski, Kramer, Kupfer, Landry, Laher, Lee, Lin, Lin, Lunnan, Giomi, Mahabal, Mao, Miller, Monkewitz, Murphy, Ngeow, Nordin, Nugent, Ofek, Patterson, Penprase, Porter, Rauch, Rebbapragada, Reiley, Rigault, Rodriguez, van Roestel, Rusholme, van Santen, Schulze, Shupe, Singer, Soumagnac, Stein, Surace, Sollerman, Szkody, Taddia, Terek, Van~Sistine, {van Velzen}, Vestrand, Walters, Ward, Ye, Yu, Yan, \& Zolkower}]{ZTF2019a}
Bellm, E.~C., Kulkarni, S.~R., Graham, M.~J., {et~al.} 2019, PASP, 131, 018002, \dodoi{10.1088/1538-3873/aaecbe}

\bibitem[{{Bode} {et~al.}(2012){Bode}, {Bogdanovi{\'c}}, {Haas}, {Healy}, {Laguna}, \& {Shoemaker}}]{Bode2012}
{Bode}, T., {Bogdanovi{\'c}}, T., {Haas}, R., {et~al.} 2012, \apj, 744, 45, \dodoi{10.1088/0004-637X/744/1/45}

\bibitem[{{Bode} {et~al.}(2010){Bode}, {Haas}, {Bogdanovi{\'c}}, {Laguna}, \& {Shoemaker}}]{Bode2010}
{Bode}, T., {Haas}, R., {Bogdanovi{\'c}}, T., {Laguna}, P., \& {Shoemaker}, D. 2010, \apj, 715, 1117, \dodoi{10.1088/0004-637X/715/2/1117}

\bibitem[{{Bogdanovi{\'c}} {et~al.}(2011){Bogdanovi{\'c}}, {Bode}, {Haas}, {Laguna}, \& {Shoemaker}}]{Bogdanovic2011}
{Bogdanovi{\'c}}, T., {Bode}, T., {Haas}, R., {Laguna}, P., \& {Shoemaker}, D. 2011, Classical and Quantum Gravity, 28, 094020, \dodoi{10.1088/0264-9381/28/9/094020}

\bibitem[{{Bolzonella} {et~al.}(2000){Bolzonella}, {Miralles}, \& {Pell{\'o}}}]{Bolzonella2000}
{Bolzonella}, M., {Miralles}, J.~M., \& {Pell{\'o}}, R. 2000, \aap, 363, 476, \dodoi{10.48550/arXiv.astro-ph/0003380}

\bibitem[{{Carrasco-Davis} {et~al.}(2021){Carrasco-Davis}, {Reyes}, {Valenzuela}, {F{\"o}rster}, {Est{\'e}vez}, {Pignata}, {Bauer}, {Reyes}, {S{\'a}nchez-S{\'a}ez}, {Cabrera-Vives}, {Eyheramendy}, {Catelan}, {Arredondo}, {Castillo-Navarrete}, {Rodr{\'\i}guez-Mancini}, {Ruz-Mieres}, {Moya}, {Sabatini-Gacit{\'u}a}, {Sep{\'u}lveda-Cobo}, {Mahabal}, {Silva-Farf{\'a}n}, {Camacho-I{\~n}iguez}, \& {Galbany}}]{Carrasco-Davis2021}
{Carrasco-Davis}, R., {Reyes}, E., {Valenzuela}, C., {et~al.} 2021, \aj, 162, 231, \dodoi{10.3847/1538-3881/ac0ef1}

\bibitem[{{Carrasco Kind} \& {Brunner}(2013)}]{CarrascoKind2013}
{Carrasco Kind}, M., \& {Brunner}, R.~J. 2013, \mnras, 432, 1483, \dodoi{10.1093/mnras/stt574}

\bibitem[{{Chandrasekhar}(1943)}]{Chandrasekhar1943}
{Chandrasekhar}, S. 1943, \apj, 97, 255, \dodoi{10.1086/144517}

\bibitem[{{CHIME/FRB Collaboration} {et~al.}(2018){CHIME/FRB Collaboration}, {Amiri}, {Bandura}, {Berger}, {Bhardwaj}, {Boyce}, {Boyle}, {Brar}, {Burhanpurkar}, {Chawla}, {Chowdhury}, {Cliche}, {Cranmer}, {Cubranic}, {Deng}, {Denman}, {Dobbs}, {Fandino}, {Fonseca}, {Gaensler}, {Giri}, {Gilbert}, {Good}, {Guliani}, {Halpern}, {Hinshaw}, {H{\"o}fer}, {Josephy}, {Kaspi}, {Landecker}, {Lang}, {Liao}, {Masui}, {Mena-Parra}, {Naidu}, {Newburgh}, {Ng}, {Patel}, {Pen}, {Pinsonneault-Marotte}, {Pleunis}, {Rafiei Ravandi}, {Ransom}, {Renard}, {Scholz}, {Sigurdson}, {Siegel}, {Smith}, {Stairs}, {Tendulkar}, {Vanderlinde}, \& {Wiebe}}]{CHIME/FRBCollaboration2018}
{CHIME/FRB Collaboration}, {Amiri}, M., {Bandura}, K., {et~al.} 2018, \apj, 863, 48, \dodoi{10.3847/1538-4357/aad188}

\bibitem[{Colpi {et~al.}(2024)Colpi, Danzmann, Hewitson, {Holley-Bockelmann}, Jetzer, Nelemans, Petiteau, Shoemaker, Sopuerta, Stebbins, Tanvir, Ward, Weber, Thorpe, Daurskikh, Deep, Fern{\'a}ndez~N{\'u}{\~n}ez, Garc{\'i}a~Marirrodriga, Gehler, Halain, Jennrich, Lammers, Larra{\~n}aga, Lieser, L{\"u}tzgendorf, Martens, Mondin, Piris~Ni{\~n}o, {Amaro-Seoane}, Arca~Sedda, Auclair, Babak, Baghi, Baibhav, Baker, Bayle, Berry, Berti, Boileau, Bonetti, Brito, Buscicchio, Calcagni, Capelo, Caprini, Caputo, Castelli, Chen, Chen, Chua, Davies, Derdzinski, Domcke, Doneva, Dvorkin, Mar{\'i}a~Ezquiaga, Gair, Haiman, Harry, Hartwig, Hees, Heffernan, Husa, {Izquierdo-Villalba}, Karnesis, Klein, Korol, Korsakova, Kupfer, Laghi, Lamberts, Larson, Le~Jeune, Lewicki, Littenberg, Madge, Mangiagli, Marsat, Vilchez, Maselli, Mathews, {van de Meent}, Muratore, Nardini, Pani, Peloso, Pieroni, Pound, {Quelquejay-Leclere}, Ricciardone, Rossi, Sartirana, Savalle, Sberna, Sesana, Shoemaker, Slutsky, Sotiriou, Speri, Staab, Steer,
  Tamanini, Tasinato, Torrado, {Torres-Orjuela}, Toubiana, Vallisneri, Vecchio, Volonteri, Yagi, \& Zwick}]{colpi2024}
Colpi, M., Danzmann, K., Hewitson, M., {et~al.} 2024, {{LISA Definition Study Report}}, \dodoi{10.48550/arXiv.2402.07571}

\bibitem[{{Connolly} {et~al.}(1995){Connolly}, {Csabai}, {Szalay}, {Koo}, {Kron}, \& {Munn}}]{Connolly1995}
{Connolly}, A.~J., {Csabai}, I., {Szalay}, A.~S., {et~al.} 1995, \aj, 110, 2655, \dodoi{10.1086/117720}

\bibitem[{{Corrales} {et~al.}(2010){Corrales}, {Haiman}, \& {MacFadyen}}]{Corrales2010}
{Corrales}, L.~R., {Haiman}, Z., \& {MacFadyen}, A. 2010, \mnras, 404, 947, \dodoi{10.1111/j.1365-2966.2010.16324.x}

\bibitem[{{Cuadra} {et~al.}(2009){Cuadra}, {Armitage}, {Alexander}, \& {Begelman}}]{Cuadra2009}
{Cuadra}, J., {Armitage}, P.~J., {Alexander}, R.~D., \& {Begelman}, M.~C. 2009, \mnras, 393, 1423, \dodoi{10.1111/j.1365-2966.2008.14147.x}

\bibitem[{{DESI Collaboration} {et~al.}(2016){DESI Collaboration}, Aghamousa, Aguilar, Ahlen, Alam, Allen, Prieto, Annis, Bailey, Balland, Ballester, Baltay, Beaufore, Bebek, Beers, Bell, Bernal, Besuner, Beutler, Blake, Bleuler, Blomqvist, Blum, Bolton, Briceno, Brooks, Brownstein, {Buckley-Geer}, Burden, Burtin, Busca, Cahn, Cai, {Cardiel-Sas}, Carlberg, Carton, Casas, Castander, {Cervantes-Cota}, Claybaugh, Close, Coker, Cole, Comparat, Cooper, Cousinou, Crocce, Cuby, Cunningham, Davis, Dawson, {de la Macorra}, De~Vicente, Delubac, Derwent, Dey, Dhungana, Ding, Doel, Duan, Ealet, Edelstein, Eftekharzadeh, Eisenstein, Elliott, Escoffier, Evatt, Fagrelius, Fan, Fanning, Farahi, Farihi, Favole, Feng, Fernandez, Findlay, Finkbeiner, Fitzpatrick, Flaugher, Flender, {Font-Ribera}, {Forero-Romero}, Fosalba, Frenk, Fumagalli, Gaensicke, Gallo, {Garcia-Bellido}, Gaztanaga, Fusillo, Gerard, Gershkovich, Giannantonio, Gillet, {Gonzalez-de-Rivera}, {Gonzalez-Perez}, Gott, Graur, Gutierrez, Guy, Habib, Heetderks,
  Heetderks, Heitmann, Hellwing, Herrera, Ho, Holland, Honscheid, Huff, Hutchinson, Huterer, Hwang, Laguna, Ishikawa, Jacobs, Jeffrey, Jelinsky, Jennings, Jiang, Jimenez, Johnson, Joyce, Jullo, Juneau, Kama, Karcher, Karkar, Kehoe, Kennamer, Kent, Kilbinger, Kim, Kirkby, Kisner, Kitanidis, Kneib, Koposov, Kovacs, Koyama, Kremin, Kron, Kronig, {Kueter-Young}, Lacey, Lafever, Lahav, Lambert, Lampton, Landriau, Lang, Lauer, Goff, Guillou, Van~Suu, Lee, Lee, Leitner, Lesser, Levi, L'Huillier, Li, Liang, Lin, Linder, Loebman, Luki{\'c}, Ma, MacCrann, Magneville, Makarem, Manera, Manser, Marshall, Martini, Massey, Matheson, McCauley, McDonald, McGreer, Meisner, Metcalfe, Miller, Miquel, Moustakas, Myers, Naik, Newman, Nichol, Nicola, {da Costa}, Nie, Niz, Norberg, Nord, Norman, Nugent, O'Brien, Oh, Olsen, Padilla, Padmanabhan, Padmanabhan, {Palanque-Delabrouille}, Palmese, Pappalardo, P{\^a}ris, Park, Patej, Peacock, Peiris, Peng, Percival, Perruchot, Pieri, Pogge, Pollack, Poppett, Prada, Prakash, Probst,
  Rabinowitz, Raichoor, Ree, Refregier, Regal, Reid, Reil, Rezaie, Rockosi, Roe, Ronayette, Roodman, Ross, Ross, Rossi, Rozo, {Ruhlmann-Kleider}, Rykoff, Sabiu, Samushia, Sanchez, Sanchez, Schlegel, Schneider, Schubnell, Secroun, Seljak, Seo, Serrano, Shafieloo, Shan, Sharples, Sholl, Shourt, Silber, Silva, Sirk, Slosar, Smith, Smoot, Som, Song, Sprayberry, Staten, Stefanik, Tarle, Tie, Tinker, Tojeiro, Valdes, Valenzuela, Valluri, {Vargas-Magana}, Verde, Walker, Wang, Wang, Weaver, Weaverdyck, Wechsler, Weinberg, White, Yang, Yeche, Zhang, Zhao, Zheng, Zhou, Zhou, Zhu, Zou, \& Zu}]{desicollaboration2016}
{DESI Collaboration}, Aghamousa, A., Aguilar, J., {et~al.} 2016, arXiv:1611.00036 [astro-ph].
\newblock \doarXiv{1611.00036}

\bibitem[{{Dilday} {et~al.}(2008){Dilday}, {Kessler}, {Frieman}, {Holtzman}, {Marriner}, {Miknaitis}, {Nichol}, {Romani}, {Sako}, {Bassett}, {Becker}, {Cinabro}, {DeJongh}, {Depoy}, {Doi}, {Garnavich}, {Hogan}, {Jha}, {Konishi}, {Lampeitl}, {Marshall}, {McGinnis}, {Prieto}, {Riess}, {Richmond}, {Schneider}, {Smith}, {Takanashi}, {Tokita}, {van der Heyden}, {Yasuda}, {Zheng}, {Barentine}, {Brewington}, {Choi}, {Crotts}, {Dembicky}, {Harvanek}, {Im}, {Ketzeback}, {Kleinman}, {Krzesi{\'n}ski}, {Long}, {Malanushenko}, {Malanushenko}, {McMillan}, {Nitta}, {Pan}, {Saurage}, {Snedden}, {Watters}, {Wheeler}, \& {York}}]{Dilday2008}
{Dilday}, B., {Kessler}, R., {Frieman}, J.~A., {et~al.} 2008, \apj, 682, 262, \dodoi{10.1086/587733}

\bibitem[{{Dotti} {et~al.}(2006){Dotti}, {Colpi}, \& {Haardt}}]{Dotti2006}
{Dotti}, M., {Colpi}, M., \& {Haardt}, F. 2006, \mnras, 367, 103, \dodoi{10.1111/j.1365-2966.2005.09956.x}

\bibitem[{{Duffell} {et~al.}(2020){Duffell}, {D'Orazio}, {Derdzinski}, {Haiman}, {MacFadyen}, {Rosen}, \& {Zrake}}]{Duffell2020}
{Duffell}, P.~C., {D'Orazio}, D., {Derdzinski}, A., {et~al.} 2020, \apj, 901, 25, \dodoi{10.3847/1538-4357/abab95}

\bibitem[{{Escala} {et~al.}(2004){Escala}, {Larson}, {Coppi}, \& {Mardones}}]{Escala2004}
{Escala}, A., {Larson}, R.~B., {Coppi}, P.~S., \& {Mardones}, D. 2004, \apj, 607, 765, \dodoi{10.1086/386278}

\bibitem[{{Fakhouri} {et~al.}(2010){Fakhouri}, {Ma}, \& {Boylan-Kolchin}}]{Fakhouri2010}
{Fakhouri}, O., {Ma}, C.-P., \& {Boylan-Kolchin}, M. 2010, \mnras, 406, 2267, \dodoi{10.1111/j.1365-2966.2010.16859.x}

\bibitem[{{Farris} {et~al.}(2010){Farris}, {Liu}, \& {Shapiro}}]{Farris2010}
{Farris}, B.~D., {Liu}, Y.~T., \& {Shapiro}, S.~L. 2010, \prd, 81, 084008, \dodoi{10.1103/PhysRevD.81.084008}

\bibitem[{{Foley} {et~al.}(2013){Foley}, {Challis}, {Chornock}, {Ganeshalingam}, {Li}, {Marion}, {Morrell}, {Pignata}, {Stritzinger}, {Silverman}, {Wang}, {Anderson}, {Filippenko}, {Freedman}, {Hamuy}, {Jha}, {Kirshner}, {McCully}, {Persson}, {Phillips}, {Reichart}, \& {Soderberg}}]{Foley2013}
{Foley}, R.~J., {Challis}, P.~J., {Chornock}, R., {et~al.} 2013, \apj, 767, 57, \dodoi{10.1088/0004-637X/767/1/57}

\bibitem[{{Fong} {et~al.}(2015){Fong}, {Berger}, {Margutti}, \& {Zauderer}}]{Fong2015}
{Fong}, W., {Berger}, E., {Margutti}, R., \& {Zauderer}, B.~A. 2015, \apj, 815, 102, \dodoi{10.1088/0004-637X/815/2/102}

\bibitem[{{F{\"o}rster} {et~al.}(2021){F{\"o}rster}, {Cabrera-Vives}, {Castillo-Navarrete}, {Est{\'e}vez}, {S{\'a}nchez-S{\'a}ez}, {Arredondo}, {Bauer}, {Carrasco-Davis}, {Catelan}, {Elorrieta}, {Eyheramendy}, {Huijse}, {Pignata}, {Reyes}, {Reyes}, {Rodr{\'\i}guez-Mancini}, {Ruz-Mieres}, {Valenzuela}, {{\'A}lvarez-Maldonado}, {Astorga}, {Borissova}, {Clocchiatti}, {De Cicco}, {Donoso-Oliva}, {Hern{\'a}ndez-Garc{\'\i}a}, {Graham}, {Jord{\'a}n}, {Kurtev}, {Mahabal}, {Maureira}, {Mu{\~n}oz-Arancibia}, {Molina-Ferreiro}, {Moya}, {Palma}, {P{\'e}rez-Carrasco}, {Protopapas}, {Romero}, {Sabatini-Gacitua}, {S{\'a}nchez}, {San Mart{\'\i}n}, {Sep{\'u}lveda-Cobo}, {Vera}, \& {Vergara}}]{Forster2021}
{F{\"o}rster}, F., {Cabrera-Vives}, G., {Castillo-Navarrete}, E., {et~al.} 2021, \aj, 161, 242, \dodoi{10.3847/1538-3881/abe9bc}

\bibitem[{{Gagliano} {et~al.}(2023){Gagliano}, {Contardo}, {Foreman-Mackey}, {Malz}, \& {Aleo}}]{Gagliano2023}
{Gagliano}, A., {Contardo}, G., {Foreman-Mackey}, D., {Malz}, A.~I., \& {Aleo}, P.~D. 2023, \apj, 954, 6, \dodoi{10.3847/1538-4357/ace326}

\bibitem[{{Gagliano} {et~al.}(2021){Gagliano}, {Narayan}, {Engel}, {Carrasco Kind}, \& {LSST Dark Energy Science Collaboration}}]{Gagliano2021}
{Gagliano}, A., {Narayan}, G., {Engel}, A., {Carrasco Kind}, M., \& {LSST Dark Energy Science Collaboration}. 2021, \apj, 908, 170, \dodoi{10.3847/1538-4357/abd02b}

\bibitem[{{Gehrels} {et~al.}(2004){Gehrels}, {Chincarini}, {Giommi}, {Mason}, {Nousek}, {Wells}, {White}, {Barthelmy}, {Burrows}, {Cominsky}, {Hurley}, {Marshall}, {M{\'e}sz{\'a}ros}, {Roming}, {Angelini}, {Barbier}, {Belloni}, {Campana}, {Caraveo}, {Chester}, {Citterio}, {Cline}, {Cropper}, {Cummings}, {Dean}, {Feigelson}, {Fenimore}, {Frail}, {Fruchter}, {Garmire}, {Gendreau}, {Ghisellini}, {Greiner}, {Hill}, {Hunsberger}, {Krimm}, {Kulkarni}, {Kumar}, {Lebrun}, {Lloyd-Ronning}, {Markwardt}, {Mattson}, {Mushotzky}, {Norris}, {Osborne}, {Paczynski}, {Palmer}, {Park}, {Parsons}, {Paul}, {Rees}, {Reynolds}, {Rhoads}, {Sasseen}, {Schaefer}, {Short}, {Smale}, {Smith}, {Stella}, {Tagliaferri}, {Takahashi}, {Tashiro}, {Townsley}, {Tueller}, {Turner}, {Vietri}, {Voges}, {Ward}, {Willingale}, {Zerbi}, \& {Zhang}}]{Gehrels2004}
{Gehrels}, N., {Chincarini}, G., {Giommi}, P., {et~al.} 2004, \apj, 611, 1005, \dodoi{10.1086/422091}

\bibitem[{{Gomez} {et~al.}(2020){Gomez}, {Berger}, {Blanchard}, {Hosseinzadeh}, {Nicholl}, {Villar}, \& {Yin}}]{Gomez2020}
{Gomez}, S., {Berger}, E., {Blanchard}, P.~K., {et~al.} 2020, \apj, 904, 74, \dodoi{10.3847/1538-4357/abbf49}

\bibitem[{Graham {et~al.}(2019)Graham, Kulkarni, Bellm, Adams, Barbarino, Blagorodnova, Bodewits, Bolin, Brady, Cenko, Chang, Coughlin, De, Eadie, Farnham, Feindt, Franckowiak, Fremling, Gezari, Ghosh, Goldstein, Golkhou, Goobar, Ho, Huppenkothen, Ivezi{\'c}, Jones, Juric, Kaplan, Kasliwal, Kelley, Kupfer, Lee, Lin, Lunnan, Mahabal, Miller, Ngeow, Nugent, Ofek, Prince, Rauch, van Roestel, Schulze, Singer, Sollerman, Taddia, Yan, Ye, Yu, Barlow, Bauer, Beck, Belicki, Biswas, Brinnel, Brooke, Bue, Bulla, Burruss, Connolly, Cromer, Cunningham, Dekany, Delacroix, Desai, Duev, Feeney, Flynn, Frederick, {Gal-Yam}, Giomi, Groom, Hacopians, Hale, Helou, Henning, Hover, Hillenbrand, Howell, Hung, Imel, Ip, Jackson, Kaspi, Kaye, Kowalski, Kramer, Kuhn, Landry, Laher, Mao, Masci, Monkewitz, Murphy, Nordin, Patterson, Penprase, Porter, Rebbapragada, Reiley, Riddle, Rigault, Rodriguez, Rusholme, van Santen, Shupe, Smith, Soumagnac, Stein, Surace, Szkody, Terek, Sistine, van Velzen, Vestrand, Walters, Ward, Zhang, \&
  Zolkower}]{ZTF2019b}
Graham, M.~J., Kulkarni, S.~R., Bellm, E.~C., {et~al.} 2019, PASP, 131, 078001, \dodoi{10.1088/1538-3873/ab006c}

\bibitem[{{Granot} {et~al.}(2002){Granot}, {Panaitescu}, {Kumar}, \& {Woosley}}]{Granot2002}
{Granot}, J., {Panaitescu}, A., {Kumar}, P., \& {Woosley}, S.~E. 2002, \apjl, 570, L61, \dodoi{10.1086/340991}

\bibitem[{{Greene} \& {Ho}(2007)}]{Greene2007}
{Greene}, J.~E., \& {Ho}, L.~C. 2007, \apj, 667, 131, \dodoi{10.1086/520497}

\bibitem[{{Hamuy} {et~al.}(1996){Hamuy}, {Phillips}, {Suntzeff}, {Schommer}, {Maza}, \& {Aviles}}]{Hamuy1996}
{Hamuy}, M., {Phillips}, M.~M., {Suntzeff}, N.~B., {et~al.} 1996, \aj, 112, 2391, \dodoi{10.1086/118190}

\bibitem[{Harris {et~al.}(2020)Harris, Millman, {van der Walt}, Gommers, Virtanen, Cournapeau, Wieser, Taylor, Berg, Smith, Kern, Picus, Hoyer, {van Kerkwijk}, Brett, Haldane, {del R{\'i}o}, Wiebe, Peterson, {G{\'e}rard-Marchant}, Sheppard, Reddy, Weckesser, Abbasi, Gohlke, \& Oliphant}]{numpy2020}
Harris, C.~R., Millman, K.~J., {van der Walt}, S.~J., {et~al.} 2020, Natur, 585, 357, \dodoi{10.1038/s41586-020-2649-2}

\bibitem[{{Holz} \& {Hughes}(2005)}]{Holz2005}
{Holz}, D.~E., \& {Hughes}, S.~A. 2005, \apj, 629, 15, \dodoi{10.1086/431341}

\bibitem[{{Hopkins} {et~al.}(2008){Hopkins}, {Hernquist}, {Cox}, \& {Kere{\v{s}}}}]{Hopkins2008}
{Hopkins}, P.~F., {Hernquist}, L., {Cox}, T.~J., \& {Kere{\v{s}}}, D. 2008, \apjs, 175, 356, \dodoi{10.1086/524362}

\bibitem[{{Hounsell} {et~al.}(2018){Hounsell}, {Scolnic}, {Foley}, {Kessler}, {Miranda}, {Avelino}, {Bohlin}, {Filippenko}, {Frieman}, {Jha}, {Kelly}, {Kirshner}, {Mandel}, {Rest}, {Riess}, {Rodney}, \& {Strolger}}]{Hounsell2018}
{Hounsell}, R., {Scolnic}, D., {Foley}, R.~J., {et~al.} 2018, \apj, 867, 23, \dodoi{10.3847/1538-4357/aac08b}

\bibitem[{Hunter(2007)}]{matplotlib2007}
Hunter, J.~D. 2007, Comput. Sci. Eng., 9, 90, \dodoi{10.1109/MCSE.2007.55}

\bibitem[{Ivezi{\'c} {et~al.}(2019)Ivezi{\'c}, Kahn, Tyson, Abel, Acosta, Allsman, Alonso, AlSayyad, Anderson, Andrew, P.~Angel, Angeli, Ansari, Antilogus, Araujo, Armstrong, Arndt, Astier, Aubourg, Auza, Axelrod, Bard, Barr, Barrau, Bartlett, Bauer, Bauman, Baumont, Bechtol, Bechtol, Becker, Becla, Beldica, Bellavia, Bianco, Biswas, Blanc, Blazek, Blandford, Bloom, Bogart, Bond, Booth, Borgland, Borne, Bosch, Boutigny, Brackett, Bradshaw, Brandt, Brown, Bullock, Burchat, Burke, Cagnoli, Calabrese, Callahan, Callen, Carlin, Carlson, Chandrasekharan, {Charles-Emerson}, Chesley, Cheu, Chiang, Chiang, Chirino, Chow, Ciardi, Claver, {Cohen-Tanugi}, Cockrum, Coles, Connolly, Cook, Cooray, Covey, Cribbs, Cui, Cutri, Daly, Daniel, Daruich, Daubard, Daues, Dawson, Delgado, Dellapenna, de~Peyster, de~{Val-Borro}, Digel, Doherty, Dubois, {Dubois-Felsmann}, Durech, Economou, Eifler, Eracleous, Emmons, Neto, Ferguson, Figueroa, {Fisher-Levine}, Focke, Foss, Frank, Freemon, Gangler, Gawiser, Geary, Gee, Geha, Gessner,
  Gibson, Gilmore, Glanzman, Glick, Goldina, Goldstein, Goodenow, Graham, Gressler, Gris, Guy, Guyonnet, Haller, Harris, Hascall, Haupt, Hernandez, Herrmann, Hileman, Hoblitt, Hodgson, Hogan, Howard, Huang, Huffer, Ingraham, Innes, Jacoby, Jain, Jammes, Jee, Jenness, Jernigan, Jevremovi{\'c}, Johns, Johnson, Johnson, Jones, {Juramy-Gilles}, Juri{\'c}, Kalirai, Kallivayalil, Kalmbach, Kantor, Karst, Kasliwal, Kelly, Kessler, Kinnison, Kirkby, Knox, Kotov, Krabbendam, Krughoff, Kub{\'a}nek, Kuczewski, Kulkarni, Ku, Kurita, Lage, Lambert, Lange, Langton, Guillou, Levine, Liang, Lim, Lintott, Long, Lopez, Lotz, Lupton, Lust, MacArthur, Mahabal, Mandelbaum, Markiewicz, Marsh, Marshall, Marshall, May, McKercher, McQueen, Meyers, Migliore, Miller, Mills, Miraval, Moeyens, Moolekamp, Monet, Moniez, Monkewitz, Montgomery, Morrison, Mueller, Muller, Arancibia, Neill, Newbry, Nief, Nomerotski, Nordby, O'Connor, Oliver, Olivier, Olsen, O'Mullane, Ortiz, Osier, Owen, Pain, Palecek, Parejko, Parsons, Pease, Peterson,
  Peterson, Petravick, Petrick, Petry, Pierfederici, Pietrowicz, Pike, Pinto, Plante, Plate, Plutchak, Price, Prouza, Radeka, Rajagopal, Rasmussen, Regnault, Reil, Reiss, Reuter, Ridgway, Riot, Ritz, Robinson, Roby, Roodman, Rosing, Roucelle, Rumore, Russo, Saha, Sassolas, Schalk, Schellart, Schindler, Schmidt, Schneider, Schneider, Schoening, Schumacher, Schwamb, Sebag, Selvy, Sembroski, Seppala, Serio, Serrano, Shaw, Shipsey, Sick, Silvestri, Slater, Smith, Smith, Sobhani, Soldahl, {Storrie-Lombardi}, Stover, Strauss, Street, Stubbs, Sullivan, Sweeney, Swinbank, Szalay, Takacs, Tether, Thaler, Thayer, Thomas, Thornton, Thukral, Tice, Trilling, Turri, Berg, Berk, Vetter, Virieux, Vucina, Wahl, Walkowicz, Walsh, Walter, Wang, Wang, Warner, Wiecha, Willman, Winters, Wittman, Wolff, {Wood-Vasey}, Wu, Xin, Yoachim, \& Zhan}]{ivezic2019}
Ivezi{\'c}, {\v Z}., Kahn, S.~M., Tyson, J.~A., {et~al.} 2019, ApJ, 873, 111, \dodoi{10.3847/1538-4357/ab042c}

\bibitem[{{Kelly} {et~al.}(2017){Kelly}, {Baker}, {Etienne}, {Giacomazzo}, \& {Schnittman}}]{Kelly2017}
{Kelly}, B.~J., {Baker}, J.~G., {Etienne}, Z.~B., {Giacomazzo}, B., \& {Schnittman}, J. 2017, \prd, 96, 123003, \dodoi{10.1103/PhysRevD.96.123003}

\bibitem[{{Kessler} {et~al.}(2010){Kessler}, {Cinabro}, {Bassett}, {Dilday}, {Frieman}, {Garnavich}, {Jha}, {Marriner}, {Nichol}, {Sako}, {Smith}, {Bernstein}, {Bizyaev}, {Goobar}, {Kuhlmann}, {Schneider}, \& {Stritzinger}}]{Kessler2010}
{Kessler}, R., {Cinabro}, D., {Bassett}, B., {et~al.} 2010, \apj, 717, 40, \dodoi{10.1088/0004-637X/717/1/40}

\bibitem[{{Khan} {et~al.}(2013){Khan}, {Holley-Bockelmann}, {Berczik}, \& {Just}}]{Khan2013}
{Khan}, F.~M., {Holley-Bockelmann}, K., {Berczik}, P., \& {Just}, A. 2013, \apj, 773, 100, \dodoi{10.1088/0004-637X/773/2/100}

\bibitem[{{Khan} {et~al.}(2020){Khan}, {Mirza}, \& {Holley-Bockelmann}}]{Khan2020}
{Khan}, F.~M., {Mirza}, M.~A., \& {Holley-Bockelmann}, K. 2020, \mnras, 492, 256, \dodoi{10.1093/mnras/stz3360}

\bibitem[{{Klein} {et~al.}(2014){Klein}, {Cornish}, \& {Yunes}}]{Klein2014}
{Klein}, A., {Cornish}, N., \& {Yunes}, N. 2014, \prd, 90, 124029, \dodoi{10.1103/PhysRevD.90.124029}

\bibitem[{{Kochanek}(2016)}]{Kochanek2016}
{Kochanek}, C.~S. 2016, \mnras, 461, 371, \dodoi{10.1093/mnras/stw1290}

\bibitem[{{Kormendy} \& {Ho}(2013)}]{Kormendy2013}
{Kormendy}, J., \& {Ho}, L.~C. 2013, \araa, 51, 511, \dodoi{10.1146/annurev-astro-082708-101811}

\bibitem[{Lacy {et~al.}(2020)Lacy, Surace, Farrah, Nyland, Afonso, Brandt, Clements, Lagos, Maraston, Pforr, Sajina, Sako, Vaccari, Wilson, Ballantyne, Barkhouse, Brunner, Cane, Clarke, Cooper, Cooray, Covone, D'Andrea, Evrard, Ferguson, Frieman, {Gonzalez-Perez}, Gupta, Hatziminaoglou, Huang, Jagannathan, Jarvis, Jones, Kimball, Lidman, Lubin, Marchetti, Martini, McMahon, Mei, Messias, Murphy, Newman, Nichol, Norris, Oliver, {Perez-Fournon}, Peters, Pierre, Polisensky, Richards, Ridgway, R{\"o}ttgering, Seymour, Shirley, Somerville, Strauss, Suntzeff, Thorman, {van~Kampen}, Verma, Wechsler, \& {Wood-Vasey}}]{Lacy2020}
Lacy, M., Surace, J.~A., Farrah, D., {et~al.} 2020, MNRAS, 501, 892, \dodoi{10.1093/mnras/staa3714}

\bibitem[{{Laghi} {et~al.}(2021){Laghi}, {Tamanini}, {Del Pozzo}, {Sesana}, {Gair}, {Babak}, \& {Izquierdo-Villalba}}]{Laghi2021}
{Laghi}, D., {Tamanini}, N., {Del Pozzo}, W., {et~al.} 2021, \mnras, 508, 4512, \dodoi{10.1093/mnras/stab2741}

\bibitem[{{Law} {et~al.}(2017){Law}, {Abruzzo}, {Bassa}, {Bower}, {Burke-Spolaor}, {Butler}, {Cantwell}, {Carey}, {Chatterjee}, {Cordes}, {Demorest}, {Dowell}, {Fender}, {Gourdji}, {Grainge}, {Hessels}, {Hickish}, {Kaspi}, {Lazio}, {McLaughlin}, {Michilli}, {Mooley}, {Perrott}, {Ransom}, {Razavi-Ghods}, {Rupen}, {Scaife}, {Scott}, {Scholz}, {Seymour}, {Spitler}, {Stovall}, {Tendulkar}, {Titterington}, {Wharton}, \& {Williams}}]{Law2017}
{Law}, C.~J., {Abruzzo}, M.~W., {Bassa}, C.~G., {et~al.} 2017, \apj, 850, 76, \dodoi{10.3847/1538-4357/aa9700}

\bibitem[{{Lippai} {et~al.}(2008){Lippai}, {Frei}, \& {Haiman}}]{Lippai2008}
{Lippai}, Z., {Frei}, Z., \& {Haiman}, Z. 2008, \apjl, 676, L5, \dodoi{10.1086/587034}

\bibitem[{{Loh} \& {Spillar}(1986)}]{Loh1986}
{Loh}, E.~D., \& {Spillar}, E.~J. 1986, \apj, 303, 154, \dodoi{10.1086/164062}

\bibitem[{{London} {et~al.}(2018){London}, {Khan}, {Fauchon-Jones}, {Garc{\'\i}a}, {Hannam}, {Husa}, {Jim{\'e}nez-Forteza}, {Kalaghatgi}, {Ohme}, \& {Pannarale}}]{London2018}
{London}, L., {Khan}, S., {Fauchon-Jones}, E., {et~al.} 2018, \prl, 120, 161102, \dodoi{10.1103/PhysRevLett.120.161102}

\bibitem[{{Lou} {et~al.}(2016){Lou}, {Liang}, {Yao}, {Zheng}, {Cheng}, {Wang}, {Liu}, {Qian}, {Zhao}, \& {Yang}}]{Lou2016}
{Lou}, Z., {Liang}, M., {Yao}, D., {et~al.} 2016, in Society of Photo-Optical Instrumentation Engineers (SPIE) Conference Series, Vol. 10154, Society of Photo-Optical Instrumentation Engineers (SPIE) Conference Series, 101542A, \dodoi{10.1117/12.2248371}

\bibitem[{{LSST Science Collaboration} {et~al.}(2009){LSST Science Collaboration}, {Abell}, {Allison}, {Anderson}, {Andrew}, {Angel}, {Armus}, {Arnett}, {Asztalos}, {Axelrod}, \& et~al.}]{LSSTScienceCollaboration2009}
{LSST Science Collaboration}, {Abell}, P.~A., {Allison}, J., {et~al.} 2009, arXiv e-prints, arXiv:0912.0201, \dodoi{10.48550/arXiv.0912.0201}

\bibitem[{{Madau} \& {Dickinson}(2014)}]{Madau2014}
{Madau}, P., \& {Dickinson}, M. 2014, \araa, 52, 415, \dodoi{10.1146/annurev-astro-081811-125615}

\bibitem[{{Mangiagli} {et~al.}(2023){Mangiagli}, {Caprini}, {Marsat}, {Speri}, {Caldwell}, \& {Tamanini}}]{Mangiagli2023}
{Mangiagli}, A., {Caprini}, C., {Marsat}, S., {et~al.} 2023, arXiv e-prints, arXiv:2312.04632, \dodoi{10.48550/arXiv.2312.04632}

\bibitem[{Mangiagli {et~al.}(2020)Mangiagli, Klein, Bonetti, Katz, Sesana, Volonteri, Colpi, Marsat, \& Babak}]{mangiagli2020}
Mangiagli, A., Klein, A., Bonetti, M., {et~al.} 2020, Phys. Rev. D, 102, 084056, \dodoi{10.1103/PhysRevD.102.084056}

\bibitem[{{Mart{\'\i}nez-Galarza} {et~al.}(2021){Mart{\'\i}nez-Galarza}, {Bianco}, {Crake}, {Tirumala}, {Mahabal}, {Graham}, \& {Giles}}]{Martinez-Galarza2021}
{Mart{\'\i}nez-Galarza}, J.~R., {Bianco}, F.~B., {Crake}, D., {et~al.} 2021, \mnras, 508, 5734, \dodoi{10.1093/mnras/stab2588}

\bibitem[{{Matheson} {et~al.}(2021){Matheson}, {Stubens}, {Wolf}, {Lee}, {Narayan}, {Saha}, {Scott}, {Soraisam}, {Bolton}, {Hauger}, {Silva}, {Kececioglu}, {Scheidegger}, {Snodgrass}, {Aleo}, {Evans-Jacquez}, {Singh}, {Wang}, {Yang}, \& {Zhao}}]{Matheson2021}
{Matheson}, T., {Stubens}, C., {Wolf}, N., {et~al.} 2021, \aj, 161, 107, \dodoi{10.3847/1538-3881/abd703}

\bibitem[{{McClintock} {et~al.}(2011){McClintock}, {Narayan}, {Davis}, {Gou}, {Kulkarni}, {Orosz}, {Penna}, {Remillard}, \& {Steiner}}]{McClintock2011}
{McClintock}, J.~E., {Narayan}, R., {Davis}, S.~W., {et~al.} 2011, Classical and Quantum Gravity, 28, 114009, \dodoi{10.1088/0264-9381/28/11/114009}

\bibitem[{McKinney(2010)}]{pandas2010}
McKinney, W. 2010, in Python in {{Science Conference}}, Austin, Texas, 56--61, \dodoi{10.25080/Majora-92bf1922-00a}

\bibitem[{{M{\"o}ller} {et~al.}(2021){M{\"o}ller}, {Peloton}, {Ishida}, {Arnault}, {Bachelet}, {Blaineau}, {Boutigny}, {Chauhan}, {Gangler}, {Hernandez}, {Hrivnac}, {Leoni}, {Leroy}, {Moniez}, {Pateyron}, {Ramparison}, {Turpin}, {Ansari}, {Allam}, {Bajat}, {Biswas}, {Boucaud}, {Bregeon}, {Campagne}, {Cohen-Tanugi}, {Coleiro}, {Dornic}, {Fouchez}, {Godet}, {Gris}, {Karpov}, {Nebot Gomez-Moran}, {Neveu}, {Plaszczynski}, {Savchenko}, \& {Webb}}]{Moller2021}
{M{\"o}ller}, A., {Peloton}, J., {Ishida}, E. E.~O., {et~al.} 2021, \mnras, 501, 3272, \dodoi{10.1093/mnras/staa3602}

\bibitem[{{Murphy} {et~al.}(2013){Murphy}, {Chatterjee}, {Kaplan}, {Banyer}, {Bell}, {Bignall}, {Bower}, {Cameron}, {Coward}, {Cordes}, {Croft}, {Curran}, {Djorgovski}, {Farrell}, {Frail}, {Gaensler}, {Galloway}, {Gendre}, {Green}, {Hancock}, {Johnston}, {Kamble}, {Law}, {Lazio}, {Lo}, {Macquart}, {Rea}, {Rebbapragada}, {Reynolds}, {Ryder}, {Schmidt}, {Soria}, {Stairs}, {Tingay}, {Torkelsson}, {Wagstaff}, {Walker}, {Wayth}, \& {Williams}}]{Murphy2013}
{Murphy}, T., {Chatterjee}, S., {Kaplan}, D.~L., {et~al.} 2013, \pasa, 30, e006, \dodoi{10.1017/pasa.2012.006}

\bibitem[{{Muthukrishna} {et~al.}(2021){Muthukrishna}, {Mandel}, {Lochner}, {Webb}, \& {Narayan}}]{Muthukrishna2021}
{Muthukrishna}, D., {Mandel}, K.~S., {Lochner}, M., {Webb}, S., \& {Narayan}, G. 2021, arXiv e-prints, arXiv:2112.08415, \dodoi{10.48550/arXiv.2112.08415}

\bibitem[{Narayan {et~al.}(2018)Narayan, Zaidi, Soraisam, Wang, Lochner, Matheson, Saha, Yang, Zhao, Kececioglu, Scheidegger, Snodgrass, Axelrod, Jenness, Maier, Ridgway, Seaman, Evans, Singh, Taylor, Toeniskoetter, Welch, Zhu, \& {The ANTARES Collaboration}}]{Narayan2018}
Narayan, G., Zaidi, T., Soraisam, M.~D., {et~al.} 2018, ApJS, 236, 9, \dodoi{10.3847/1538-4365/aab781}

\bibitem[{{Palenzuela} {et~al.}(2010){Palenzuela}, {Lehner}, \& {Liebling}}]{Palenzuela2010}
{Palenzuela}, C., {Lehner}, L., \& {Liebling}, S.~L. 2010, Science, 329, 927, \dodoi{10.1126/science.1191766}

\bibitem[{{Perez-Carrasco} {et~al.}(2023){Perez-Carrasco}, {Cabrera-Vives}, {Hernandez-Garc{\'\i}a}, {F{\"o}rster}, {Sanchez-Saez}, {Mu{\~n}oz Arancibia}, {Arredondo}, {Astorga}, {Bauer}, {Bayo}, {Catelan}, {Dastidar}, {Est{\'e}vez}, {Lira}, \& {Pignata}}]{Perez-Carrasco2023}
{Perez-Carrasco}, M., {Cabrera-Vives}, G., {Hernandez-Garc{\'\i}a}, L., {et~al.} 2023, \aj, 166, 151, \dodoi{10.3847/1538-3881/ace0c1}

\bibitem[{Peters(1964)}]{Peters1964}
Peters, P.~C. 1964, Phys. Rev., 136, B1224, \dodoi{10.1103/PhysRev.136.B1224}

\bibitem[{{Peterson}(2014)}]{Peterson2014}
{Peterson}, B.~M. 2014, \ssr, 183, 253, \dodoi{10.1007/s11214-013-9987-4}

\bibitem[{{Petroff} {et~al.}(2019){Petroff}, {Hessels}, \& {Lorimer}}]{Petroff2019}
{Petroff}, E., {Hessels}, J.~W.~T., \& {Lorimer}, D.~R. 2019, \aapr, 27, 4, \dodoi{10.1007/s00159-019-0116-6}

\bibitem[{{Phillips}(1993)}]{Phillips1993}
{Phillips}, M.~M. 1993, \apjl, 413, L105, \dodoi{10.1086/186970}

\bibitem[{{Predehl} {et~al.}(2021){Predehl}, {Andritschke}, {Arefiev}, {Babyshkin}, {Batanov}, {Becker}, {B{\"o}hringer}, {Bogomolov}, {Boller}, {Borm}, {Bornemann}, {Br{\"a}uninger}, {Br{\"u}ggen}, {Brunner}, {Brusa}, {Bulbul}, {Buntov}, {Burwitz}, {Burkert}, {Clerc}, {Churazov}, {Coutinho}, {Dauser}, {Dennerl}, {Doroshenko}, {Eder}, {Emberger}, {Eraerds}, {Finoguenov}, {Freyberg}, {Friedrich}, {Friedrich}, {F{\"u}rmetz}, {Georgakakis}, {Gilfanov}, {Granato}, {Grossberger}, {Gueguen}, {Gureev}, {Haberl}, {H{\"a}lker}, {Hartner}, {Hasinger}, {Huber}, {Ji}, {Kienlin}, {Kink}, {Korotkov}, {Kreykenbohm}, {Lamer}, {Lomakin}, {Lapshov}, {Liu}, {Maitra}, {Meidinger}, {Menz}, {Merloni}, {Mernik}, {Mican}, {Mohr}, {M{\"u}ller}, {Nandra}, {Nazarov}, {Pacaud}, {Pavlinsky}, {Perinati}, {Pfeffermann}, {Pietschner}, {Ramos-Ceja}, {Rau}, {Reiffers}, {Reiprich}, {Robrade}, {Salvato}, {Sanders}, {Santangelo}, {Sasaki}, {Scheuerle}, {Schmid}, {Schmitt}, {Schwope}, {Shirshakov}, {Steinmetz}, {Stewart}, {Str{\"u}der},
  {Sunyaev}, {Tenzer}, {Tiedemann}, {Tr{\"u}mper}, {Voron}, {Weber}, {Wilms}, \& {Yaroshenko}}]{Predehl2021}
{Predehl}, P., {Andritschke}, R., {Arefiev}, V., {et~al.} 2021, \aap, 647, A1, \dodoi{10.1051/0004-6361/202039313}

\bibitem[{{Qu} \& {Sako}(2023)}]{Qu2023}
{Qu}, H., \& {Sako}, M. 2023, \apj, 954, 201, \dodoi{10.3847/1538-4357/aceafa}

\bibitem[{{Quinlan}(1996)}]{Quinlan1996}
{Quinlan}, G.~D. 1996, \na, 1, 35, \dodoi{10.1016/S1384-1076(96)00003-6}

\bibitem[{{Ravi}(2018)}]{Ravi2018}
{Ravi}, V. 2018, in Astronomical Society of the Pacific Conference Series, Vol. 517, Science with a Next Generation Very Large Array, ed. E.~{Murphy}, 781, \dodoi{10.48550/arXiv.1806.08446}

\bibitem[{{Rees}(1988)}]{Rees1988}
{Rees}, M.~J. 1988, \nat, 333, 523, \dodoi{10.1038/333523a0}

\bibitem[{{Rodriguez-Gomez} {et~al.}(2015){Rodriguez-Gomez}, {Genel}, {Vogelsberger}, {Sijacki}, {Pillepich}, {Sales}, {Torrey}, {Snyder}, {Nelson}, {Springel}, {Ma}, \& {Hernquist}}]{Rodriguez-Gomez2015}
{Rodriguez-Gomez}, V., {Genel}, S., {Vogelsberger}, M., {et~al.} 2015, \mnras, 449, 49, \dodoi{10.1093/mnras/stv264}

\bibitem[{{Rossi} {et~al.}(2010){Rossi}, {Lodato}, {Armitage}, {Pringle}, \& {King}}]{Rossi2010}
{Rossi}, E.~M., {Lodato}, G., {Armitage}, P.~J., {Pringle}, J.~E., \& {King}, A.~R. 2010, \mnras, 401, 2021, \dodoi{10.1111/j.1365-2966.2009.15802.x}

\bibitem[{Sako {et~al.}(2008)Sako, Bassett, Becker, Cinabro, DeJongh, Depoy, Dilday, Doi, Frieman, Garnavich, Hogan, Holtzman, Jha, Kessler, Konishi, Lampeitl, Marriner, Miknaitis, Nichol, Prieto, Riess, Richmond, Romani, Schneider, Smith, SubbaRao, Takanashi, Tokita, van~der Heyden, Yasuda, Zheng, Barentine, Brewington, Choi, Dembicky, Harnavek, Ihara, Im, Ketzeback, Kleinman, Krzesi{\'n}ski, Long, Malanushenko, Malanushenko, McMillan, Morokuma, Nitta, Pan, Saurage, \& Snedden}]{sako2008}
Sako, M., Bassett, B., Becker, A., {et~al.} 2008, AJ, 135, 348, \dodoi{10.1088/0004-6256/135/1/348}

\bibitem[{{Santamar{\'\i}a} {et~al.}(2010){Santamar{\'\i}a}, {Ohme}, {Ajith}, {Br{\"u}gmann}, {Dorband}, {Hannam}, {Husa}, {M{\"o}sta}, {Pollney}, {Reisswig}, {Robinson}, {Seiler}, \& {Krishnan}}]{Santamaria2010}
{Santamar{\'\i}a}, L., {Ohme}, F., {Ajith}, P., {et~al.} 2010, \prd, 82, 064016, \dodoi{10.1103/PhysRevD.82.064016}

\bibitem[{{Schnittman} \& {Krolik}(2008)}]{Schnittman2008}
{Schnittman}, J.~D., \& {Krolik}, J.~H. 2008, \apj, 684, 835, \dodoi{10.1086/590363}

\bibitem[{{Sesana} {et~al.}(2004){Sesana}, {Haardt}, {Madau}, \& {Volonteri}}]{Sesana2004}
{Sesana}, A., {Haardt}, F., {Madau}, P., \& {Volonteri}, M. 2004, \apj, 611, 623, \dodoi{10.1086/422185}

\bibitem[{{Sesana} \& {Khan}(2015)}]{Sesana2015}
{Sesana}, A., \& {Khan}, F.~M. 2015, \mnras, 454, L66, \dodoi{10.1093/mnrasl/slv131}

\bibitem[{{Smartt}(2009)}]{Smartt2009}
{Smartt}, S.~J. 2009, \araa, 47, 63, \dodoi{10.1146/annurev-astro-082708-101737}

\bibitem[{Spergel {et~al.}(2015)Spergel, Gehrels, Baltay, Bennett, Breckinridge, Donahue, Dressler, Gaudi, Greene, Guyon, Hirata, Kalirai, Kasdin, Macintosh, Moos, Perlmutter, Postman, Rauscher, Rhodes, Wang, Weinberg, Benford, Hudson, Jeong, Mellier, Traub, Yamada, Capak, Colbert, Masters, Penny, Savransky, Stern, Zimmerman, Barry, Bartusek, Carpenter, Cheng, Content, Dekens, Demers, Grady, Jackson, Kuan, Kruk, Melton, Nemati, Parvin, Poberezhskiy, Peddie, Ruffa, Wallace, Whipple, Wollack, \& Zhao}]{WFIRST2015}
Spergel, D., Gehrels, N., Baltay, C., {et~al.} 2015, arXiv:1503.03757 [astro-ph].
\newblock \doarXiv{1503.03757}

\bibitem[{{Strolger} {et~al.}(2015){Strolger}, {Dahlen}, {Rodney}, {Graur}, {Riess}, {McCully}, {Ravindranath}, {Mobasher}, \& {Shahady}}]{Strolger2015}
{Strolger}, L.-G., {Dahlen}, T., {Rodney}, S.~A., {et~al.} 2015, \apj, 813, 93, \dodoi{10.1088/0004-637X/813/2/93}

\bibitem[{{Sullivan} {et~al.}(2006){Sullivan}, {Howell}, {Perrett}, {Nugent}, {Astier}, {Aubourg}, {Balam}, {Basa}, {Carlberg}, {Conley}, {Fabbro}, {Fouchez}, {Guy}, {Hook}, {Lafoux}, {Neill}, {Pain}, {Palanque-Delabrouille}, {Pritchet}, {Regnault}, {Rich}, {Taillet}, {Aldering}, {Baumont}, {Bronder}, {Filiol}, {Knop}, {Perlmutter}, \& {Tao}}]{Sullivan2006}
{Sullivan}, M., {Howell}, D.~A., {Perrett}, K., {et~al.} 2006, \aj, 131, 960, \dodoi{10.1086/499302}

\bibitem[{{Tamanini}(2017)}]{Tamanini2017}
{Tamanini}, N. 2017, in Journal of Physics Conference Series, Vol. 840, Journal of Physics Conference Series (IOP), 012029, \dodoi{10.1088/1742-6596/840/1/012029}

\bibitem[{The Astropy~Collaboration {et~al.}(2013)The Astropy~Collaboration, Robitaille, Tollerud, Greenfield, Droettboom, Bray, Aldcroft, Davis, Ginsburg, {Price-Whelan}, Kerzendorf, Conley, Crighton, Barbary, Muna, Ferguson, Grollier, Parikh, Nair, G{\"u}nther, Deil, Woillez, Conseil, Kramer, Turner, Singer, Fox, Weaver, Zabalza, Edwards, Bostroem, Burke, Casey, Crawford, Dencheva, Ely, Jenness, Labrie, Lim, Pierfederici, Pontzen, Ptak, Refsdal, Servillat, \& Streicher}]{astropy2013}
The Astropy~Collaboration, A., Robitaille, T.~P., Tollerud, E.~J., {et~al.} 2013, A\&A, 558, \dodoi{10.1051/0004-6361/201322068}

\bibitem[{{Thornton} {et~al.}(2013){Thornton}, {Stappers}, {Bailes}, {Barsdell}, {Bates}, {Bhat}, {Burgay}, {Burke-Spolaor}, {Champion}, {Coster}, {D'Amico}, {Jameson}, {Johnston}, {Keith}, {Kramer}, {Levin}, {Milia}, {Ng}, {Possenti}, \& {van Straten}}]{Thornton2013}
{Thornton}, D., {Stappers}, B., {Bailes}, M., {et~al.} 2013, Science, 341, 53, \dodoi{10.1126/science.1236789}

\bibitem[{Tuttle {et~al.}(2017)Tuttle, Vaillon, Johann, Wallner, \& Ergenzinger}]{Euclid2017}
Tuttle, S., Vaillon, L., Johann, U., Wallner, O., \& Ergenzinger, K. 2017, in International {{Conference}} on {{Space Optics}} --- {{ICSO}} 2010, ed. N.~Kadowaki (Rhodes Island, Greece: SPIE), 90, \dodoi{10.1117/12.2309226}

\bibitem[{{van Paradijs} {et~al.}(2000){van Paradijs}, {Kouveliotou}, \& {Wijers}}]{vanParadijs2000}
{van Paradijs}, J., {Kouveliotou}, C., \& {Wijers}, R. A.~M.~J. 2000, \araa, 38, 379, \dodoi{10.1146/annurev.astro.38.1.379}

\bibitem[{{van Velzen}(2018)}]{vanVelzen2018}
{van Velzen}, S. 2018, \apj, 852, 72, \dodoi{10.3847/1538-4357/aa998e}

\bibitem[{{Villar} {et~al.}(2017){Villar}, {Berger}, {Metzger}, \& {Guillochon}}]{Villar2017}
{Villar}, V.~A., {Berger}, E., {Metzger}, B.~D., \& {Guillochon}, J. 2017, \apj, 849, 70, \dodoi{10.3847/1538-4357/aa8fcb}

\bibitem[{{Villar} {et~al.}(2021){Villar}, {Cranmer}, {Berger}, {Contardo}, {Ho}, {Hosseinzadeh}, \& {Lin}}]{Villar2021}
{Villar}, V.~A., {Cranmer}, M., {Berger}, E., {et~al.} 2021, \apjs, 255, 24, \dodoi{10.3847/1538-4365/ac0893}

\bibitem[{Virtanen {et~al.}(2020)Virtanen, Gommers, Oliphant, Haberland, Reddy, Cournapeau, Burovski, Peterson, Weckesser, Bright, {van der Walt}, Brett, Wilson, Millman, Mayorov, Nelson, Jones, Kern, Larson, Carey, Polat, Feng, Moore, VanderPlas, Laxalde, Perktold, Cimrman, Henriksen, Quintero, Harris, Archibald, Ribeiro, Pedregosa, {van Mulbregt}, {SciPy 1.0 Contributors}, Vijaykumar, Bardelli, Rothberg, Hilboll, Kloeckner, Scopatz, Lee, Rokem, Woods, Fulton, Masson, H{\"a}ggstr{\"o}m, Fitzgerald, Nicholson, Hagen, Pasechnik, Olivetti, Martin, Wieser, Silva, Lenders, Wilhelm, Young, Price, Ingold, Allen, Lee, Audren, Probst, Dietrich, Silterra, Webber, Slavi{\v c}, Nothman, Buchner, Kulick, Sch{\"o}nberger, {de Miranda Cardoso}, Reimer, Harrington, Rodr{\'i}guez, {Nunez-Iglesias}, Kuczynski, Tritz, Thoma, Newville, K{\"u}mmerer, Bolingbroke, Tartre, Pak, Smith, Nowaczyk, Shebanov, Pavlyk, Brodtkorb, Lee, McGibbon, Feldbauer, Lewis, Tygier, Sievert, Vigna, Peterson, More, Pudlik, Oshima, Pingel, Robitaille,
  Spura, Jones, Cera, Leslie, Zito, Krauss, Upadhyay, Halchenko, \& {V{\'a}zquez-Baeza}}]{scipy2020}
Virtanen, P., Gommers, R., Oliphant, T.~E., {et~al.} 2020, NatMe, 17, 261, \dodoi{10.1038/s41592-019-0686-2}

\bibitem[{{Volonteri}(2010)}]{Volonteri2010}
{Volonteri}, M. 2010, \aapr, 18, 279, \dodoi{10.1007/s00159-010-0029-x}

\bibitem[{{Volonteri} {et~al.}(2003){Volonteri}, {Haardt}, \& {Madau}}]{Volonteri2003}
{Volonteri}, M., {Haardt}, F., \& {Madau}, P. 2003, \apj, 582, 559, \dodoi{10.1086/344675}

\bibitem[{{Volonteri} {et~al.}(2020){Volonteri}, {Pfister}, {Beckmann}, {Dubois}, {Colpi}, {Conselice}, {Dotti}, {Martin}, {Jackson}, {Kraljic}, {Pichon}, {Trebitsch}, {Yi}, {Devriendt}, \& {Peirani}}]{Volonteri2020}
{Volonteri}, M., {Pfister}, H., {Beckmann}, R.~S., {et~al.} 2020, \mnras, 498, 2219, \dodoi{10.1093/mnras/staa2384}

\bibitem[{{Wanderman} \& {Piran}(2010)}]{Wanderman2010}
{Wanderman}, D., \& {Piran}, T. 2010, \mnras, 406, 1944, \dodoi{10.1111/j.1365-2966.2010.16787.x}

\bibitem[{{Wyithe} \& {Loeb}(2003)}]{Wyithe2003}
{Wyithe}, J. S.~B., \& {Loeb}, A. 2003, \apj, 590, 691, \dodoi{10.1086/375187}

\bibitem[{{Yao} {et~al.}(2023){Yao}, {Ravi}, {Gezari}, {van Velzen}, {Lu}, {Schulze}, {Somalwar}, {Kulkarni}, {Hammerstein}, {Nicholl}, {Graham}, {Perley}, {Cenko}, {Stein}, {Ricarte}, {Chadayammuri}, {Quataert}, {Bellm}, {Bloom}, {Dekany}, {Drake}, {Groom}, {Mahabal}, {Prince}, {Riddle}, {Rusholme}, {Sharma}, {Sollerman}, \& {Yan}}]{Yao2023}
{Yao}, Y., {Ravi}, V., {Gezari}, S., {et~al.} 2023, \apjl, 955, L6, \dodoi{10.3847/2041-8213/acf216}

\bibitem[{{Yuan} {et~al.}(2021){Yuan}, {Murase}, {Zhang}, {Kimura}, \& {M{\'e}sz{\'a}ros}}]{Yuan2021}
{Yuan}, C., {Murase}, K., {Zhang}, B.~T., {Kimura}, S.~S., \& {M{\'e}sz{\'a}ros}, P. 2021, \apjl, 911, L15, \dodoi{10.3847/2041-8213/abee24}

\bibitem[{{Zanotti} {et~al.}(2010){Zanotti}, {Rezzolla}, {Del Zanna}, \& {Palenzuela}}]{Zanotti2010}
{Zanotti}, O., {Rezzolla}, L., {Del Zanna}, L., \& {Palenzuela}, C. 2010, \aap, 523, A8, \dodoi{10.1051/0004-6361/201014969}

\bibitem[{{Zhan}(2018)}]{Zhan2018}
{Zhan}, H. 2018, in 42nd COSPAR Scientific Assembly, Vol.~42, E1.16--4--18

\end{thebibliography}
\bibliographystyle{aasjournal}

\end{CJK*}
\end{document}